\documentclass[twocolumn,times, trackchanges]{aastex63}
\usepackage{cleveref}
\usepackage{enumerate}
\usepackage{enumitem}
\shorttitle{HST/WFC3 Reference Star Differential Imaging of the PDS 70 Planetary System}
\shortauthors{Sanghi, A., Zhou, Y., \& Bowler, B., P.}

\begin{document}

\title{\textbf{Efficiently Imaging Accreting Protoplanets from Space: \\Reference Star Differential Imaging of the PDS 70 Planetary System using the HST/WFC3 Archival PSF Library}}

\correspondingauthor{Aniket Sanghi}
\email{asanghi01@utexas.edu}

\author[0000-0002-1838-4757]{Aniket Sanghi}
\affiliation{The University of Texas at Austin, Department of Astronomy, 2515 Speedway, C1400, Austin, TX 78712, USA}

\author[0000-0003-2969-6040]{Yifan Zhou}
\altaffiliation{51 Pegasi b Fellow}
\affiliation{The University of Texas at Austin, Department of Astronomy, 2515 Speedway, C1400, Austin, TX 78712, USA}
\affiliation{McDonald Observatory, Fort Davis, TX 79734, USA}

\author[0000-0003-2649-2288]{Brendan P. Bowler}
\affiliation{The University of Texas at Austin, Department of Astronomy, 2515 Speedway, C1400, Austin, TX 78712, USA}

\begin{abstract}
Accreting protoplanets provide key insights into how planets assemble from their natal protoplanetary disks. Recently, \citet{Zhou_2021} used angular differential imaging (ADI) with Hubble Space Telescope's Wide Field Camera 3 (HST/WFC3) to recover the young accreting planet PDS 70 b in F656N ($\mathrm{H}\alpha$) at a S/N of 7.9. In this paper, we demonstrate a promising approach to efficiently imaging accreting planets by applying reference star differential imaging (RDI) to the same dataset. We compile a reference library from the database of WFC3 point-spread functions (PSFs) provided by Space Telescope Science Institute and develop a set of morphology-significance criteria for pre-selection of reference frames to improve RDI subtraction. RDI with this PSF library results in a detection of PDS 70 b at a S/N of 5.3. Astrometry and photometry of PDS 70 b are calibrated using a forward-modeling method and injection-recovery tests, resulting in a separation of $186 \pm 13$ mas, a position angle of $142 \pm 5^\circ$, and an H$\alpha$ flux of $(1.7 \pm 0.3)\times10^{-15}$ erg s$^{-1}$ cm$^{-2}$. The lower detection significance with RDI can be attributed to the $\sim$100 times lower peak-to-background ratios of the reference PSFs compared to the ADI PSFs. Building a high-quality reference library with WFC3 will provide unique opportunities to study accretion variability on short timescales not limited by roll angle scheduling constraints and efficiently search for actively accreting protoplanets in $\mathrm{H}\alpha$ around targets inaccessible to ground-based adaptive optics systems, such as faint transition disk hosts.
\end{abstract}


\section{Introduction}

Over the past two decades, instruments such as high-order adaptive optics (AO) systems and coronagraphs combined with innovative observing and post-processing methods have enabled the direct detection and characterization of exoplanets \citep[e.g.,][]{2006ApJS..167...81G,2009ARA&A..47..253O, 2010Sci...329...57L,2010SPIE.7736E..1JM,2010A&ARv..18..317A,2015Sci...350...64M,2016PASP..128j2001B}. High-contrast imaging has generally probed exoplanets at separations $>$10 au and masses $\gtrsim$1 $M_\mathrm{{Jup}}$. Uncovering exoplanets at large separations has led to the exploration of different planet formation mechanisms such as core accretion, dynamical scattering, disk instability, and cloud fragmentation, all of which act over different orbital distance regimes \citep[e.g.,][]{1997Sci...276.1836B, 2014prpl.conf..643H, 2014prpl.conf..619C, 2014Life....4..142H, 2016ARA&A..54..271K}. The direct observation of photons originating in the atmospheres of young exoplanets allows for measurements of effective temperature, composition, and thermal structure \citep[e.g.,][]{2010ApJ...723..850B, 2020A&A...644A..13S, 2020AJ....159..263W, 2021A&A...646A.150O}. In recent years, the direct detections of $\mathrm{H}\alpha$ emission from newly formed planets \citep{2015Natur.527..342S, Wagner_2018, 2019NatAs...3..749H, 2020AJ....159..222H, Zhou_2021} have constrained planetary-mass accretion rates and enabled quantitative studies of accretion physics, planet-disk interactions, and planetary luminosity evolution \citep[e.g.,][]{2007ApJ...655..541M, 2015ApJ...809...93D, 2017ApJ...836..221M, 2018ApJ...866...84A, 2019ApJ...884L..41B}. All of these themes directly benefit from achieving high instrument sensitivity at smaller angular separations from the host star in order to detect and characterize long-period giant planets and accreting protoplanets. 

Space-based observatories offer several advantages compared to their ground-based counterparts in high-contrast exoplanet imaging. Space telescopes are not hindered by the turbulence in the Earth's atmosphere, which disrupts the image point spread function (PSF), degrading the achievable angular resolution. Ground-based observatories thus require AO systems to correct the incoming distorted wavefronts and reach high Strehl ratios. Presently, there are a limited number of optical AO systems in operation --- VLT/MUSE \citep{2010SPIE.7735E..08B}, MagAO/MagAO-X \citep{2014ApJ...786...32M, 2015ApJ...815..108M, 2018SPIE10703E..09M, 2020SPIE11448E..4LM}, Robo-AO \citep{2014ApJ...790L...8B}, SCExAO \citep{2015PASP..127..890J}, and SPHERE/ZIMPOL \citep{2018A&A...619A...9S}. For these systems, the brightness of the target star as a AO calibrator is a critical factor for successful AO correction. In addition, space-based observatories enable more accurate and stable photometry for absolute photometric calibration and time-resolved photometric measurements.

The Hubble Space Telescope (HST) is a powerful facility for high-contrast exoplanet and disk imaging. For example, over the past two decades, HST has been used to detect debris-disk systems that might host planets \citep{2014ApJ...786L..23S, 2014AJ....148...59S}, investigate the Fomalhaut system \citep[e.g.,][]{2008Sci...322.1345K, 2013ApJ...775...56K, 2020PNAS..117.9712G}, provide follow-up observations to the planetary mass companion 2M1207b \citep{2006ApJ...652..724S, 2019AJ....157..128Z}, and characterize the HR8799 planets \citep{Lafreni_re_2009, 2011ApJ...741...55S, 2015ApJ...809L..33R}. In particular, HST's Wide Field Camera 3 UVIS channel (WFC3/UVIS) enables diffraction-limited imaging at ultraviolet (UV) and optical wavelengths. This capability provides unique opportunities for the detection of young self-luminous giant exoplanets in the process of forming using accretion-induced H$\alpha$ emission \citep{2014ApJ...783L..17Z, 2015Natur.527..342S}. Recently, \citet{Zhou_2021} demonstrated high-contrast exoplanet imaging with WFC3/UVIS at small angular separations with their detection of PDS 70 b. 

PDS 70 is a young ($\sim$5 Myr) K7 T Tauri star in the Upper Sco association \citep{2016MNRAS.461..794P}. The star is undergoing modest accretion \citep{2020ApJ...892...81T} and hosts a disk with complex structures including a giant inner cavity \citep{2012ApJ...758L..19H, 2019A&A...625A.118K}. Ground-based high-contrast imaging observations uncovered two planets, PDS 70 b and c, within the disk cavity \citep{2018A&A...617A..44K, 2019NatAs...3..749H, 2020AJ....159..263W, 2021AJ....161..148W} located at projected separations of 20 and 34 au from the star, respectively. \citet{2019ApJ...879L..25I} tentatively detected the presence of a circumplanetary disk around PDS 70 c, which was recently confirmed by \citet{2021ApJ...916L...2B} using Atacama Large Millimeter/submillimeter Array (ALMA) observations of the dust continuum emission at 855 $\mu$m.

Presently, direct-imaging observations of the PDS 70 planets have primarily employed the angular differential imaging (ADI; \citealt{2004Sci...305.1442L, 2006ApJ...641..556M}) strategy. The implementations of ADI between space- and ground-based observations are different but follow the same principle. In space-based observations, images of the host star are obtained at different telescope roll angles. In ground-based observations, the telescope de-rotator is turned off. As the orientation of the telescope changes in the former case, and the field-of-view rotates under sidereal motion in the latter case, astrophysical signals (e.g., companions and disk features) revolve in the image frame while the point spread function of the host star remains stationary on the detector. Subtracting image frames obtained at different position angles (P.A.s) then removes the host starlight without significant subtraction of the planet signal.

While ADI has proved to be an effective observational strategy for directly imaging exoplanets, it possesses certain inherent limitations. In order to detect close-in companions without self-subtraction of the planet signal, each target must be observed with a sufficient P.A. difference --- the companion must move by a substantial fraction of the width of the PSF. In the case of HST observations, the maximum telescope rotation that can be achieved between two successive orbits is approximately 30$^\circ$. For images in the F656N band ($\lambda_{\mathrm{eff}} = 6561$ Å, FWHM$ = 17.9$ Å), if we assume a minimum separation limit of 1.5 FWHM (60 mas) between two ADI images, our inner working angle is constrained to 115 mas. Thus, for inner working angles $\lesssim$ 115 mas, ADI with WFC3/UVIS will introduce self-subtraction of point sources. Because of the limited permitted P.A. difference between adjacent orbits, scheduling constraints mean that longer observing baselines with non-contiguous orbits are needed to expand the P.A. coverage. Thus, ADI observations with HST incur a significant time cost. For example, to obtain an adequate ADI sequence for the PDS 70 dataset used in \citet{Zhou_2021} and this work, the observation campaign included six visits spanning a total of five months. Thus, if the scientific goal of a high-contrast imaging observation is to probe phenomena at the planetary rotation timescale or shorter (a few hours), ADI alone would be insufficient.

The above limitations can be mitigated with the use of reference star differential imaging (RDI; \citealt{Lafreni_re_2009}). This approach uses reference stars distinct from the science target to build a model of the target star's PSF for primary-star subtraction. The application of RDI has been limited by the technique's requirement of a large library of high-quality reference PSFs. In this paper, we leverage Space Telescope Science Institute's (STScI's) archival WFC3 PSF library to construct a set of reference images for the purpose of performing RDI on the PDS 70 observational dataset presented in \citet{Zhou_2021}. We investigate the sensitivity of imaging planets with RDI and explore the best practices in implementing RDI with HST.

This paper is organized as follows. Section \ref{observations} describes the WFC3 observations of PDS 70. Section \ref{library} details the construction of the reference star library using the WFC3 archival PSF library and the selection criteria adopted to characterize the quality of PSFs in the reference library. Section \ref{RDI} discusses the application of RDI to the observational dataset and methods to determine the calibrated astrometry and photometry for the detection of PDS 70 b. Section \ref{contrast} characterizes the signal-to-noise ratio (S/N) of the detection and details the contrast curve calculations. Section \ref{discussion} compares the RDI results with those from ADI and investigates possible reasons for the lower S/N detection of PDS 70 b with RDI. Finally, we summarize our conclusions in Section \ref{conclusion}.

\section{Observations}
\label{observations}
PDS 70 was observed with WFC3 in its
UVIS channel for 18 orbits (Program GO-15830\footnote{The detailed observing plan can be found here: \url{https://www.stsci.edu/hst/phase2-public/15830.pdf}}, PI: Zhou). All observations were conducted in the direct-imaging mode with the \texttt{UVIS2/C512C} subarray (field-of-view: $20\farcs2 \times 20\farcs2$). The observations were constructed as six visit sets, each consisting of three contiguous HST orbits. They were executed on UT dates 2020-02-07, 2020-04-08, 2020-05-07, 2020-05-08, 2020-06-19, and 2020-07-03. As part of the angular differential imaging (ADI) strategy implemented in \citet{Zhou_2021}, the telescope orientation angle was increased by at least 10 degrees from orbit to orbit. The total angular rotation of the telescope was $196\fdg2$. They used the F336W ($\lambda_{\mathrm{eff}} = 3359$ Å, FWHM$ = 550$ Å) and F656N ($\lambda_{\mathrm{eff}} = 6561$ Å, FWHM$ = 17.9$ Å) filters to measure the flux on the blue side of the Balmer jump and in the $\mathrm{H}\alpha$ line, respectively. Note that our experiments are restricted to narrow-band images in F656N to avoid complications caused by color differences between the target and the reference stars. For the F656N observations, each orbit consisted of nine 20 s exposures. Over 18 orbits, these observations amount to 162 images with a total integration time of 3240 s. More details on the observations can be found in \citet{Zhou_2021}.

\section{HST/WFC3 Archival PSF Library}
\label{library}

\subsection{Assembly of the Reference PSF Library}
In order to enable reference star differential imaging (RDI) for our PDS 70 dataset, we compiled a reference star library from the archival collection of WFC3 UVIS PSFs \citep{2021wfc..rept...12D} available on STScI's MAST portal\footnote{\url{https://mast.stsci.edu/portal/Mashup/Clients/Mast/Portal.html}}. The entire database contains more than 20 million PSFs, and includes WFC3/UVIS observations taken between 2009 and 2018. Each reference PSF has an array size of $21 \times 21$ pixels. The database can be queried using the following parameters --- $\textit{Focus}$, $\textit{Filter}$, $\textit{PSF X Center}$, $\textit{PSF Y Center}$, $\textit{PSF Flux}$, \texttt{QFIT}, $\textit{Exposure Time}$, $\textit{Aperture}$, $\textit{Chip}$, $\textit{MJD}$, and $\textit{Program ID}$. \texttt{QFIT} characterizes the quality of WFC3 PSFs by measuring the absolute value of the residuals of a template fit to the library PSF. This measurement is normalized such that \texttt{QFIT} lies between 0 and 1. For this study we use query parameters $\textit{Filter} = \textit{F656N}$, $\textit{PSF X Center} = [6, 512]$, $\textit{PSF Y Center} = [6, 512]$, and $\textit{Chip} = 2$. The selections are based on the following considerations:

1. The limits on the \emph{PSF X} and \emph{Y Center} parameters constrain the library PSFs within the same subarray -- \texttt{C512C} -- as the target frames to avoid significant geometric distortion.

2. No restrictions are placed on the focus since the five-month observational baseline of the PDS 70 dataset encompasses a wide range of focus values. Thus, restricting the focus should not provide any additional gains in the pre-selection process. Moreover, we avoid overly strict criteria to ensure the availability of a sufficient number of PSFs for application of the Karhunen–Loève Image Processing \citep[KLIP:][]{2012ApJ...755L..28S} algorithm.

3. Since additional down-selection procedures based on PSF shape and quality are conducted in the following subsection, no restrictions are placed on PSF quality-related parameters like PSF flux, exposure time, aperture, and \texttt{QFIT}.

The above conditions yielded 112 calibrated, flat-fielded reference frames for our PSF library.

\subsection{Reference PSF Selection for Primary Subtraction}
\label{selection:section}

\begin{figure*}[p]
\centering
    \includegraphics[scale=0.22]{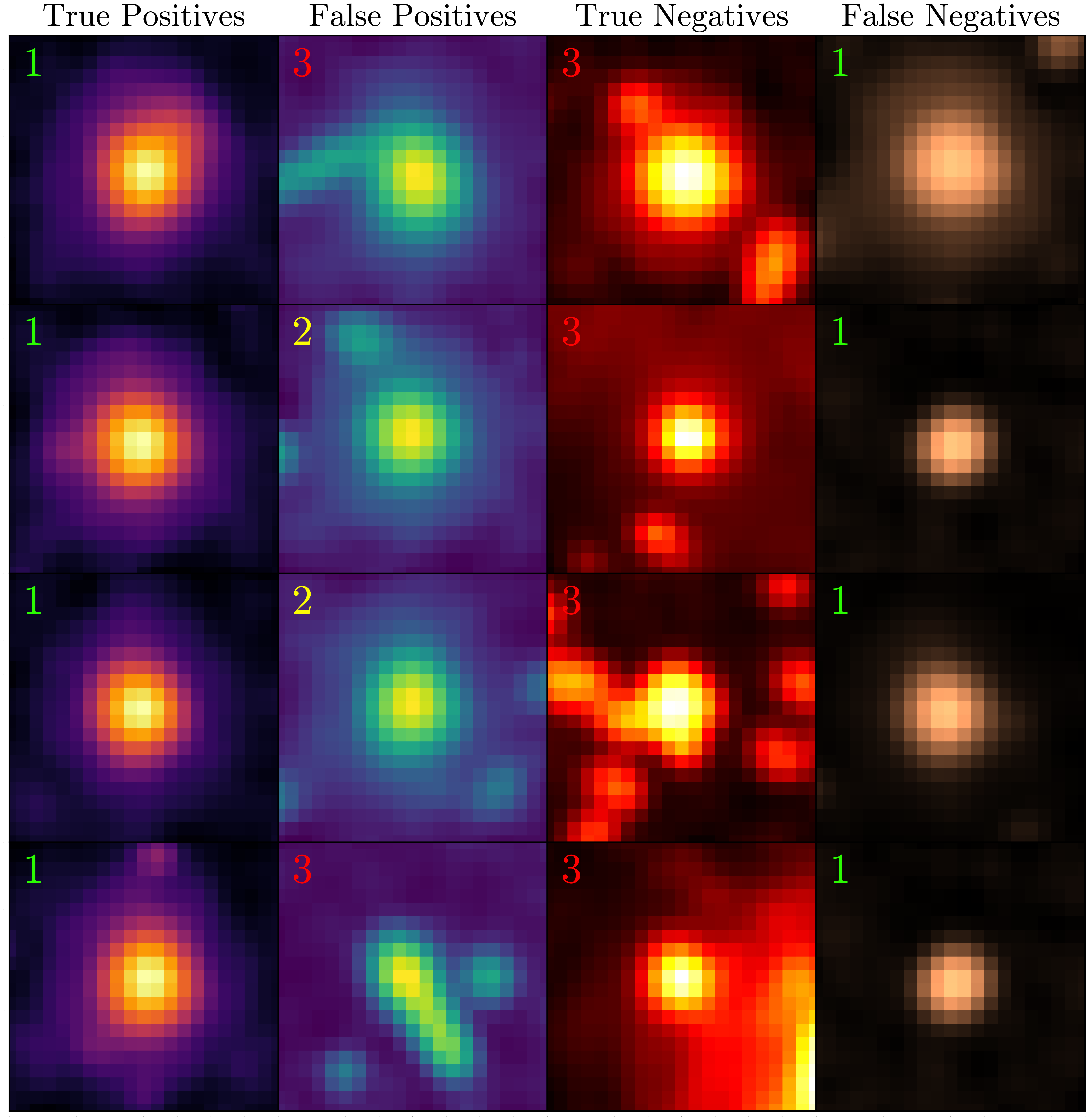}
    \caption{Example PSFs from the archival WFC3 PSF Library categorized in columns as true positives, false positives, true negatives, and false negatives based on the comparison between the visually assigned quality flag and the automated results of the morphology-significance criteria. This mosaic also provides representative examples for each quality flag category, which is represented as a number on the top left of each image frame; 1 represents good-quality PSFs, 2 represents intermediate-quality PSFs, and 3 represents bad-quality PSFs.}
    \label{fig:psfpanel}
\end{figure*}

A crucial factor in determining the performance of RDI is the construction of a good quality reference star library \citep{2019AJ....157..118R}. Reference frame pre-selection is important to achieve high contrast ratios at small angular separations, but this is not a straightforward problem. The WFC3 archival library contains PSFs that were collected across multiple epochs under different telescope conditions, observing modes, exposure times and PSF fluxes as part of several observing programs with varying levels of contamination from nearby background sources. Each of these can contribute to the final quality of the PSF, precluding the use of a single parameter to make PSF selection for RDI on target frames. PSFs that should be avoided include those with contaminating background features, extended structures, cosmic rays, or asymmetric or non-circular PSF shapes. Examples of WFC3 library PSFs representing varying PSF quality levels are presented in Figure \ref{fig:psfpanel} as a visual reference.

One solution is to visually inspect the reference library and select good quality PSFs. However, this approach works only in the case of small reference libraries and is susceptible to subjective judgements. For larger reference libraries, which are generally desired to improve primary subtraction, a preferred strategy is to develop a set of robust quantitative selection criteria that can be applied uniformly to the entire reference library to obtain high-quality reference PSFs. In this study, we leverage the small size of our reference library to first assign visually determined quality flags to each of the reference PSFs, as described below. Then, we develop a set of selection criteria that can be applied to other HST observations to construct a high-quality reference PSF library. The performance of the selection criteria is then assessed by testing it against our visually assigned quality flags.

Three quality flags are assigned based on visual inspection of PSFs in the reference library: 

1. ``Quality 1" PSFs are good-quality PSFs. They are roughly symmetric in shape and do not have contaminating sources.

2. ``Quality 2" PSFs are intermediate-quality PSFs. They are slightly asymmetric or elongated along one axis and have few faint contaminating sources.

3. ``Quality 3" PSFs are bad-quality PSFs. They are distinguished by the presence of extended structures or bundles of pixels affected by cosmic rays across the image frame, bright background sources contaminating the reference PSF, visual binaries, or severely non-circular PSF shapes.

Our selection results in 45 Quality 1 PSFs, 14 Quality 2 PSFs, and 53 Quality 3 PSFs. Example PSFs in each quality-flag category are presented in Figure \ref{fig:psfpanel}. We note that these visual inspection results serve as a baseline for the following automated PSF selection procedures; the Quality 1 PSFs are ultimately used for PSF subtraction.

Next, we develop a set of morphology-significance criteria in order to uniformly and automatically characterize the PSF quality. The morphology criterion \emph{M} quantitatively characterizes the symmetry of the PSF under consideration. It is defined as the ratio of the $1\sigma$ standard deviation along the semimajor axis of the PSF to the $1\sigma$ standard deviation along the semiminor axis of the PSF. Here, the axes are defined as the semimajor/semiminor axis of the 2D Gaussian function that has the same second-order central moments as the source PSF. An image's second-order central moments $\mu_{pq}$, $0 \le p, q \le 2$, form the elements of a $3 \times 3$ covariance matrix that is used to identify the image's major and minor axes of intensity and derive information about image orientation. We find that the image orientation is uniformly distributed for near-circular PSFs, i.e., limiting PSF image selection by orientation does not improve the quality of the library. Mathematically, the elements of the above covariance matrix are defined by the equation
\begin{equation}
    \mu_{pq} = \sum\limits_{x}\sum\limits_{y} (x -\overline{x})^p (y -\overline{y})^q I(x, y),
\end{equation}
\noindent where $I(x, y)$ represents the pixel intensities in the image and $(\overline{x}, \overline{y})$ represent the PSF centroid coordinates. In practice, we calculate the $1\sigma$ standard deviation along the semimajor and semiminor axis using the \texttt{data\_properties} function from the python \texttt{photutils.morphology} module \citep{2019zndo...3568287B}. The resulting ratio is a dimensionless quantity.

The peak significance criterion $P$ characterizes the signal significance. It is defined as the ratio of peak intensity to the median background level of the PSF cutout, and serves as an alternative to S/N. The peak-to-background ratio uses values easily accessible from the images. This helps keep the criterion simple and consistent for all the reference PSFs. For an image of dimensions $N_{\mathrm{pix}} \times N_{\mathrm{pix}}$, the peak intensity $I_p$ is calculated by averaging the peak intensities obtained after summing pixel values across each axis, normalized by the total number of pixels summed over, as follows:
\begin{equation}
    I_{p,y} = \max \left(\frac{1}{N_{\mathrm{pix}}}\sum\limits_{x} I(x, y)\right),
\end{equation}
\begin{equation}
    I_{p,x} = \max \left(\frac{1}{N_{\mathrm{pix}}}\sum\limits_{y} I(x, y)\right),
\end{equation}
\begin{equation}
    I_p = \frac{I_{p,y} + I_{p,x}}{2}.
\end{equation}
\noindent Finally, the median background level is determined using an iterative sigma clipping routine with a $3\sigma$ clipping threshold. This iterative process excludes pixels around the source centroid. The resulting ratio is a dimensionless quantity.

Sharp, symmetric PSFs with high peak significance are desirable. These are characterized by values of $M$ near 1 (a circular PSF) and large values of $P$. Quantitatively, a suitable set of selection criteria would be $1 \le M \le 1 + \delta$ and $P \ge 3$, where $\delta$ represents scatter from the ideal ratio of 1 for a symmetric PSF and $P \ge 3$ represents a signal with peak intensity at least three times greater than the computed median background level. Note that a lower bound of $1 - \delta$ is not physically meaningful, as the 1$\sigma$ semimajor-semiminor axis ratio can never be less than one. The criterion $P > 3$ excludes $\sim$36\% of our reference library. 

To determine a reasonable value of $\delta$, we consider breathing effects of HST due to thermal expansions and contractions (HST/WFC3 Manual Version 13.0\footnote{\url{https://www.stsci.edu/files/live/sites/www/files/home/hst/instrumentation/wfc3/_documents/wfc3_ihb.pdf}}). For WFC3, \citet{2012wfc..rept...14D} found that breathing induces small temporal variations in the UVIS PSF. Based on the same study, the variations of the UVIS PSF FWHM are expected to be as large as 8\% at 200 nm, 3\% at 400 nm and 0.3\% at 800 nm, on typical time scales of one orbit. These data provide the characteristic range of FWHM variations when HST is in normal operation. For our observations in the F656N filter, we approximate the variation to be $\sim$1\% of the FWHM using a quadratic fit to the above data. If $\delta$ is set to +1\% of the theoretical FWHM at 656 nm, it would exclude 60\% of the reference library and ultimately select only 23 reference PSFs. This may be insufficient for optimal RDI subtraction. Thus, we selected $\delta$ to be +3\% of the WFC3 theoretical FWHM at 656 nm ($\sim$1.7 UVIS pixels). Independently, this selection criterion for $M$ excludes $\sim$32\% of our reference library. A larger library would provide more flexibility to restrict this parameter in the future.

In addition to the above criteria, we experiment with the performance of the \texttt{QFIT} parameter provided by STScI and made available for look-up from the PSF library. \texttt{QFIT} aims to characterize the quality of WFC3 PSFs. It is defined as the absolute value of the residuals of a template fit to the library PSF’s inner $5\times5$ pixels divided by the flux of the star \citep{2012ApJ...753....7S} and ranges from 0 to 1, where 0 represents the best-quality PSFs and 1 represents the worst-quality PSFs. The template fit to the PSF is called the ``effective PSF", which is the instrumental PSF convolved with the pixel-response function of the detector \citep{2008PASP..120..907M}. In our comparisons, we select PSFs with \texttt{QFIT} $<$ 0.1. These are PSFs with fit residual intensity less than 10\% of the flux of the star.

\begin{figure}[t]
    \centering
    \includegraphics[scale=0.11]{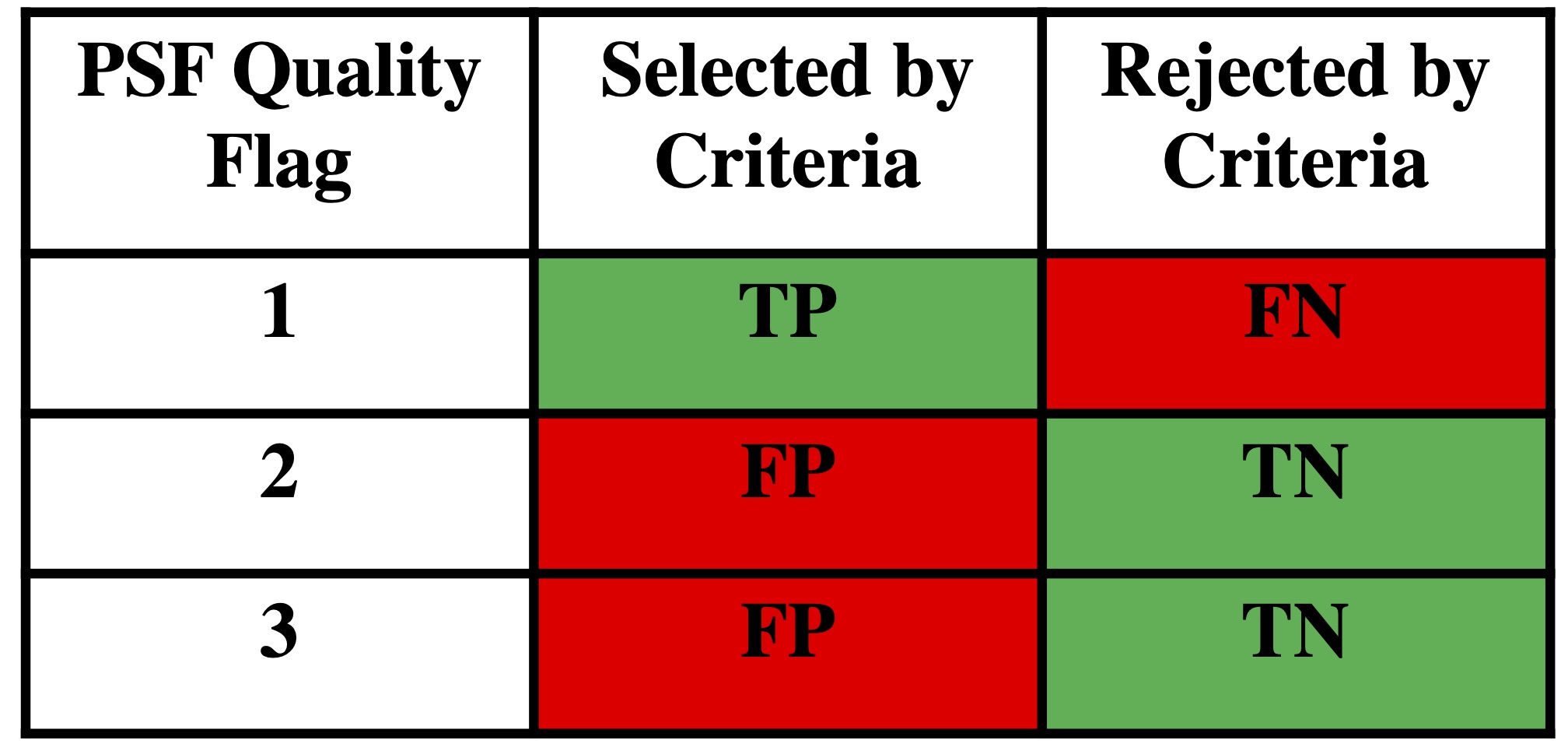}
    \caption{Classification of the reference PSFs based on the results obtained from the morphology-significance criteria, defined by the PSF's quality flag and whether it was selected or rejected by the criteria. $TP$ stands for true positive, $TN$ for true negative, $FP$ for false positive, and $FN$ for false negative. These classifications provide a way to characterize the success and failure rate of the selection criteria. Green cells highlight the expected result combinations ($TP$, $TN$) and red cells highlight the anomalous result combinations ($FP$, $FN$).}
    \label{fig:table}
\end{figure}

\begin{figure*}[t]
    \centering
    \includegraphics[scale=0.75]{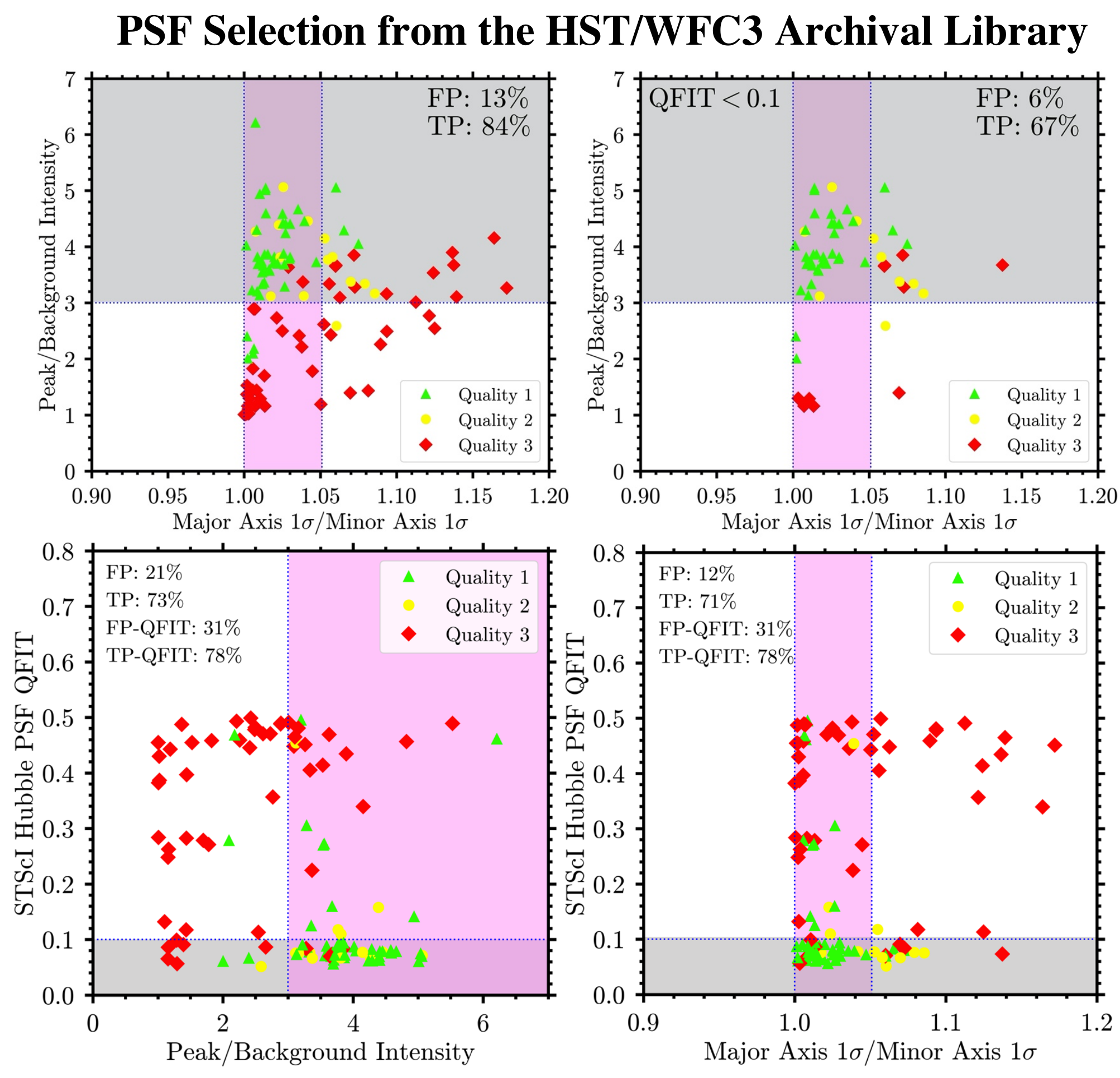}
    \caption{PSF selection using the morphology-significance criteria and STScI's \texttt{QFIT} parameter. Green triangles represent Quality 1 PSFs, yellow circles represent Quality 2 PSFs, and red diamonds represent Quality 3 PSFs. The gray and pink regions are the selection cuts for the $y$- and $x$-axis parameters, respectively, across all panels. The intersection of the gray and pink regions is the final selection region. \emph{Top left:} Selection using the morphology-significance criteria. \emph{Top right:} Selection using the combination of the morphology-significance criteria and the \texttt{QFIT} parameter. All PSFs with \texttt{QFIT} $\ge 0.1$ have been excluded from the plot and calculated $TP$ and $FP$ percentages are shown. \emph{Bottom:} These panels show the results of selecting PSFs using a combination of the \texttt{QFIT} parameter and one of the morphology-significance criteria. The calculated $TP$ and $FP$ percentages are shown for the combined application as well as for the sole application of the \texttt{QFIT} parameter ($FP$-\texttt{QFIT}, $TP$-\texttt{QFIT}).}
    \label{fig:selection}
\end{figure*}

The performance of the morphology-significance criteria is characterized using true and false positive and negative fractions. Figure \ref{fig:table} provides a succinct visual guide for which PSFs are classified as true positives ($TP$), true negatives ($TN$), false positives ($FP$), and false negatives ($FN$) based on whether they were chosen or rejected by our selection criteria. The true positive fraction ($TPF$) is defined as $TPF = \frac{TP}{TP + FN}$ and the false positive fraction ($FPF$) is defined as $FPF = \frac{FP}{FP + TN}$. The true negative fraction ($TNF$) is defined as $TNF = 1 - FPF$ and the false negative fraction ($FNF$) is defined as $FNF = 1 - TPF$. Example PSFs in each statistical category are presented in Figure \ref{fig:psfpanel}, and results after applying them to the $\mathrm{H}\alpha$ library are presented in Figure \ref{fig:selection}.

\begin{figure*}[t]
    \centering
    \includegraphics[scale=0.2]{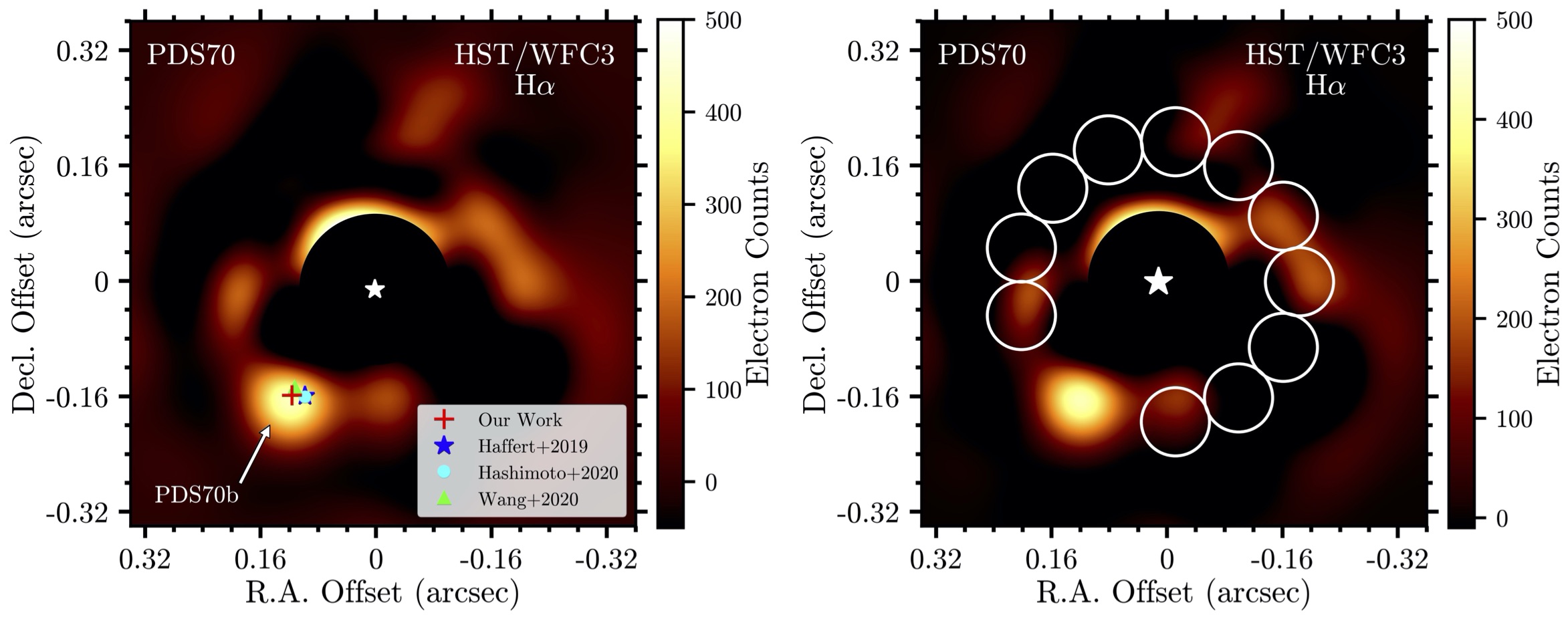}
    \caption{\emph{Left:} Final primary-subtracted image depicting our detection of PDS 70 b after performing RDI KLIP subtraction on 162 $\mathrm{H}\alpha$ images. The above images are resampled by a factor of two using 2D cubic spline interpolation. To reduce high frequency noise, we smooth these images with a Gaussian kernel with a FWHM of 2 pixels ($\sim$60\% of the PSF FWHM). North is up and east is to the left. Astrometric measurements from literature are plotted for comparison. \emph{Right:} A demonstration of how the S/N is calculated based on the methodology in \citet{2014ApJ...792...97M}, using an aperture size of 2.36 pixels ($\sim$1.4 FWHM).}
    \label{fig:RDI}
\end{figure*}

The morphology-significance criteria automatically select reference PSFs from the compiled library with a $TPF$ of 84\% and a $FPF$ of 13\%. If this is combined with STScI's \texttt{QFIT} parameter, the $TPF$ reduces to 67\% and the corresponding $FPF$ also reduces to 6\%. False positives most commonly occur due to low intensity of the contaminating features that were observed in the visual selection process but which the automatic procedure failed to identify.

The morphology-significance criteria automatically reject reference PSFs from the compiled library with a $TNF$ of 87\% and a $FNF$ of 16\%. If this is combined with STScI's \texttt{QFIT} parameter, the $TNF$ increases to 94\% and the corresponding $FNF$ also increases to 33\%. False negatives most commonly occur due to slightly higher semimajor-semiminor axis ratios or lower peak-to-background intensities that were rejected because of the strictness of our criteria limits.

Since higher true positive and negative fractions can be achieved when selecting PSFs just using the morphology-significance criteria, this alone is suitable for small reference libraries where both false positives and negatives can be filtered out by visual inspection, after the automatic selection. On the other hand, since lower false positive and negative fractions can be achieved when the morphology-significance criteria is combined with STScI’s \texttt{QFIT} parameter, the combined selection criteria are suitable for large reference libraries where it may be time-consuming to filter out the false positives and negatives by visual inspection.

\begin{figure*}[t]
    \centering
    \includegraphics[scale=0.95]{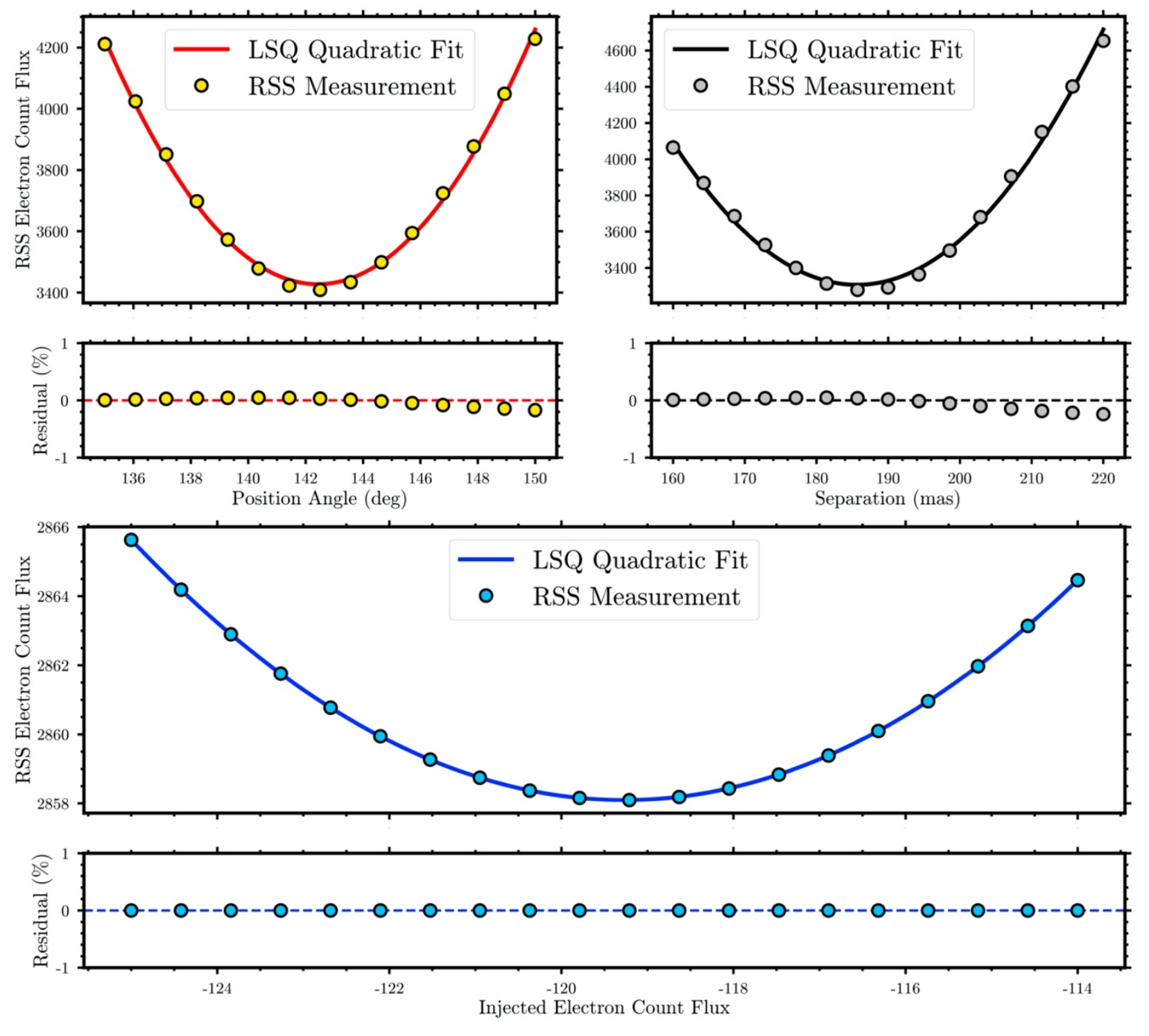}
    \caption{Calibrating astrometry and photometry of our detection of PDS 70 b with forward modeling. \emph{Top panel:} RSS electron count flux as a function of position angle and separation based on a quadratic least-squares (LSQ) fit to our measurements. Sub-panels show the fit residuals as a percentage of the RSS electron count flux for each measurement. A P.A. of $142^\circ$ with respect to north and a separation of $186$ mas from the host star minimizes the residuals. \emph{Bottom Panel:} RSS electron count flux as a function of injected electron count flux. The sub-panel shows the fit residuals as a percentage of the RSS electron count flux for each measurement. A total absolute electron count flux of 119.3 electron counts or, equivalently, $1.7\times10^{-15}$ erg s$^{-1}$ cm$^{-2}$, minimizes the residuals based on a quadratic least-squares (LSQ) fit to the measurements. These values are adopted as our final astrometry and photometry for PDS 70 b.}
    \label{fig:FM}
\end{figure*}

\begin{figure*}[t]
    \centering
    \includegraphics[scale=0.14]{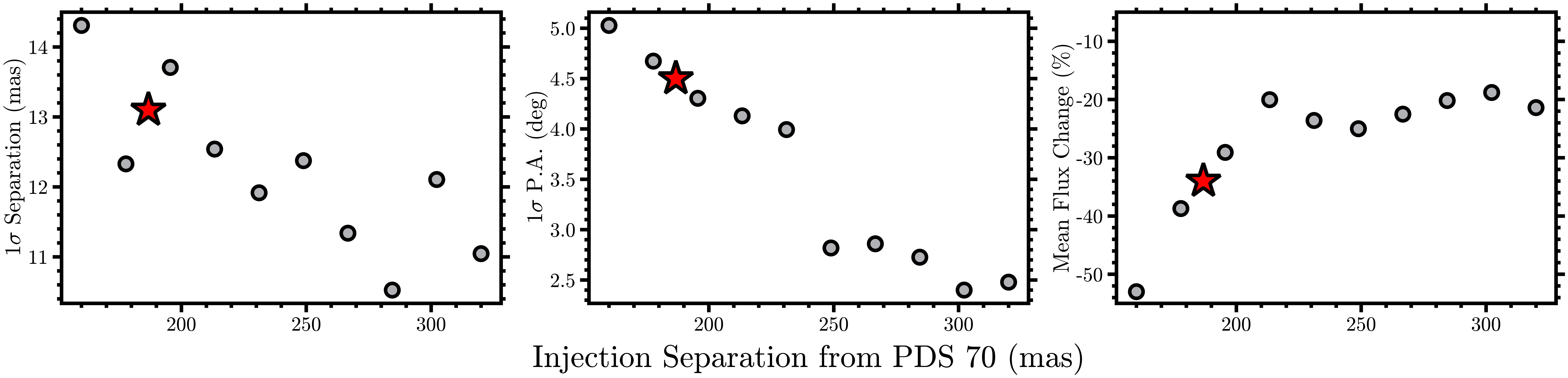}
    \caption{Error estimated by injection-recovery tests. The synthetic PSF is injected with the same positive flux as that obtained for PDS 70 b from forward modeling. Left: $1\sigma$ uncertainty in separation as a function of injection separation from PDS 70. Center: $1\sigma$ uncertainty in P.A. as a function of injection separation from PDS 70. Right: Mean flux change percentage as a function of injection separation from PDS 70. The astrometric uncertainty and mean flux change percentage at the separation of PDS 70 b (red stars) are calculated as the mean value of adjacent measurements.}
    \label{fig:IR}
\end{figure*}

\section{Reference Star Differential Imaging}
\label{RDI}
In this section, we apply reference star differential imaging to WFC3 images of PDS 70 to subtract the PSF of the host star; measure the planet's calibrated astrometry, photometry and associated uncertainties; and optimize our method to maximize the S/N of planet detections with our compiled WFC3 reference library. 

We start with the \texttt{CalWFC3} pipeline-product \texttt{flc} images of PDS 70 downloaded from the MAST archive. These files are similar to the flat-field corrected \texttt{flt} frames and have also been corrected for charge transfer efficiency losses. Our procedures include three steps: (1) primary star PSF subtraction using the KLIP algorithm; (2) forward modeling of the planet detection to calibrate astrometry and photometry; and (3) TinyTim PSF injection-recovery testing to characterize our uncertainties on the above quantities. These three steps are described below.

\subsection{High-contrast Imaging Reduction}
\label{HCI}
PSF subtraction using RDI is carried out for 162 $\mathrm{H}\alpha$ WFC3 images of PDS 70. Unlike the ADI sequence of PDS 70, the reference PSFs in the archival WFC3 library are not dithered. Thus, we cannot use the Fourier reconstruction method described in \citet{1999PASP..111..227L} to enable high-contrast imaging reduction on upsampled images \citep{Zhou_2021}. Instead, the images in this analysis have the WFC3 native pixel scale of 40 mas, which is slightly below the Nyquist criterion at $\mathrm{H}\alpha$. 

Rather than using the entire Quality flag = 1 PSF library, we instead also experiment with random subsets of the library with different numbers of images to find which combination of reference stars provides the best planet detection. This enables us to evaluate the effect of different selections of PSFs from the reference library on the corresponding detection limits and test the robustness of our results. 

For each randomly selected PSF subset, we perform a complete set of primary subtraction procedures. First, the library PSFs are registered to the science frames by determining the centroid using the \texttt{photutils} package’s \texttt{centroid\_2dg} function and implementing sub-pixel shifts using the \texttt{image\_registration} package’s \texttt{chi2\_shift} function \citep{2013A&A...558A..33A, 2018AJ....156..123A}. We then perform PSF subtraction using KLIP on all our target images. The algorithm was performed on a $21 \times 21$ pixel ($0.84"\times0.84"$) subarray centered on the PDS 70, which is determined by the array size of our reference images. Various geometries of the subtraction region are tested: 1 to 4 annular sectors centered on the host star, each constructed at radial separations varying between 2 to 9 pixels, and subdivided into 1 to 3 equal-size segments. The configuration that results in the best planet detection S/N is selected. In the optimal solution, the target image is divided into annular sectors with radial bounds at 2, 6, and 9 pixels and no angular separators subdivide the sectors. KLIP is performed independently in each of the annular sectors for each image. The number of KLIP components is also determined empirically to maximize the detection S/N. This number is always found to be equal to the number of PSFs in the selected subset of reference stars, i.e., the maximum allowable number. Finally, the PSF-subtracted images are aligned to true north, stacked, and the combined image is upsampled using a cubic spline interpolation to obtain the final post-processed image.

The best detection of PDS 70 b obtained after RDI KLIP subtraction is shown in Figure \ref{fig:RDI}. We characterize the S/N of our detection, provide a 5$\sigma$ contrast curve, and study the distribution of $5\sigma$ contrast curves obtained with different subsets of our selected reference library in Section \ref{contrast}.

\subsection{Forward Modeling the PDS 70 b Detection}
\label{fm_section}
We conduct our photometric and astrometric measurements of PDS 70 b using the KLIP forward modeling method \citep[e.g.,][]{2016ApJ...824..117P}. This technique calibrates measurement biases introduced in the primary subtraction procedures. The major steps in the procedure are: (1) simulate synthetic PSFs with varying negative intensities; (2) inject the synthetic PSFs in a grid centered on the point source being modeled; (3) perform KLIP subtraction on images containing the negative PSFs; and (4) compute the residual sum of squares flux around the point source being modeled. The principle behind forward modeling a point source detection is that when a negative intensity PSF that matches the true signal in intensity and position is injected into the original images, it cancels out the astrophysical signal and, therefore, minimizes the residuals in the primary-subtracted images. 

Synthetic WFC3 UVIS2 F656N PSFs are generated with the TinyTim package \citep{10.1117/12.892762}. We note that TinyTim only models the telescope diffraction effects and does not include the effects of any optical ghosts or halos \citep{10.1117/12.892762}. For WFC3, TinyTim uses aberration coefficients derived from the as-built optical ray trace model as a function of field position. While the results match expectations, these coefficients have not been refined based on their observed field dependence\footnote{\url{https://www.stsci.edu/hst/instrumentation/focus-and-pointing/focus/tiny-tim-hst-psf-modeling}}. Presently, models for WFC3 also do not support space-variant charge diffusion kernels \citep{2008wfc..rept...14H}. Thus, a TinyTim PSF may not represent the true HST instrumental PSF at a level sufficient for forward modeling. To test this, we compare the results obtained using the TinyTim model PSF with those using an observational template PSF (see Appendix for further details). We find that the TinyTim PSF and the observational template PSF give consistent results, validating our use of synthetic PSFs for these purposes.

We choose the following options when executing \emph{tiny1}, the routine used to generate TinyTim PSFs: \emph{Choice = 22} (WFC3 UVIS Channel), \emph{Detector = 2}, \emph{Position = (256, 256)}, \emph{Filter passband = F656N}, \emph{Diameter of PSF = 6 arcseconds}, and \emph{Focus, secondary mirror despace = 2.5 microns}. We use the PSF output with a pixel scale of 0.04 arcseconds, which is the same as our observational dataset. The orginal frame size of the simulated PSFs is $179 \times 179$ pixels. This is cropped to a $5 \times 5$ pixel frame for injection. Ideally, we would need to forward model the astrometry and photometry of the planet simultaneously. However, due to the computational complexity of carrying this out on a three-dimensional grid, we forward model astrometry and photometry separately while keeping the other fixed. In our implementation, we inject the forward model signal into the \texttt{flc} image frames before image reduction.

To measure the astrometry by forward modeling our detection, we create a position grid around the visually approximated centroid of PDS 70 b --- a P.A. of 141$^\circ$ and a separation of 185 mas. The grid is defined in terms of polar coordinates with 15 separation positions from PDS 70 (160 to 220 mas in steps of 4 mas) and 15 position angles ($135^\circ$ to $150^\circ$ from true north in steps of $1^\circ$). We fix the total PSF electron count flux to --112 counts for the negative TinyTim PSF to be injected into our images. This number was obtained by converting the flux measurement of PDS 70 b obtained by \citet{Zhou_2021} ($1.62 \pm 0.22 \times 10^{-15}$ erg s$^{-1}$ cm$^{-2}$) to electron counts. We then inject the negative simulated TinyTim PSF at each grid point into all of the 162 target images of PDS 70, perform KLIP subtraction using our reference library, and then calculate the residual sum of squares (RSS) electron count flux in a fixed $4 \times 4$ pixel aperture, centered on the visually approximated centroid of PDS 70 b. Once we have obtained the RSS electron count flux for each grid point, we marginalize the resulting 2D distribution of RSS values across each axis to obtain the distribution of RSS electron count flux as a function of separation and position angle separately. We perform a quadratic least-squares (LSQ) fit to the above distributions and locate the separation and position angle corresponding to the minimum RSS electron count flux in the fit. These quantities are designated as our forward-modeled astrometric parameters. We obtained a P.A. of $142^\circ$ with respect to true north and a separation of $186$ mas from the host star PDS 70 as the set of parameters that minimized the residuals after subtraction. Our results are depicted in the top panel of Figure \ref{fig:FM}.

To calibrate the flux of PDS 70 b with forward modeling, we follow a procedure similar to the one described above, with 20 different negative electron count flux values (--125 to --114 electron counts in steps of 0.6 electron counts) while fixing the injection location of the TinyTim PSF to the forward-modeled astrometry parameters. Upon performing a quadratic least-squares (LSQ) fit to the data, we find that the total absolute electron flux value that minimizes the residuals after subtraction is 119.3 electron counts. We convert the above count rate to flux in erg s$^{-1}$ cm$^{-2}$ using the F656N filter width and the \texttt{PHOTFLAM} inverse sensitivity factors provided in the \texttt{FITS} file headers to obtain a flux of $1.7\times10^{-15}$ erg s$^{-1}$ cm$^{-2}$. Results are presented in the bottom panel of Figure \ref{fig:FM}.

\subsection{Uncertainty in PDS 70 b Astrometry and Photometry}
\label{ir-testing}
We perform PSF injection-recovery tests using TinyTim PSFs \citep{10.1117/12.892762} to determine the uncertainty on our astrometric and photometric measurements as well as compute the KLIP throughput loss. The best-fit synthetic PSF found in Section \ref{fm_section} is injected in an annular position grid defined in terms of polar coordinates made up of 10 separation positions from PDS 70 (160 to 320 mas in steps of 18 mas) and 10 position angles ($-160^\circ$ to $90^\circ$ from true north in steps of $28^\circ$). Note that injections are performed with a minimal difference of 55$^\circ$ from the location of PDS 70 b in order to prevent contamination of the synthetic signal. We inject the synthetic PSF at each grid point across all 162 images of PDS 70, perform KLIP subtraction, and measure the astrometry and photometry of the recovered source. The centroid of the recovered source is determined by fitting a 2D Gaussian function that has the same second-order central moments as the source PSF. This centroid is used to compute the final separation and position angle. The recovered flux is estimated by performing aperture photometry using a 3 pixel aperture centered on the above calculated centroid. The result is corrected for aperture loss by determining the flux lost when an aperture of the same size is used to estimate the flux of a synthetic TinyTim PSF before injection. Finally, the error in position angle, separation, and the percentage flux lost after aperture loss correction, are all recorded. We compute the standard deviation of the errors in separation and position angle as a function of injected separation by computing statistics over results obtained across multiple position angles at the same separation. We also compute the mean flux change percentage as a function of separation to estimate throughput loss. Figure \ref{fig:IR} shows our results. To determine the astrometric uncertainty, we extract the $1\sigma$ separation and position angle at the separation of PDS 70 b (186 mas) by calculating the mean value of adjacent measurements. This results in a separation of $186 \pm 13$ mas and a position angle of $142 \pm 5^\circ$.

The photometric uncertainty consists of three components: the speckle noise, the photon noise, and the KLIP throughput uncertainty. Speckle noise is determined from the standard deviation of the flux integrated within the apertures illustrated in Figure \ref{fig:RDI}. This component also accounts for possible contamination from the circumstellar disk. Based on the ratio of the calibrated flux of PDS 70 b and the speckle noise, we determine the S/N for our detection of PDS 70 b to be 5.3. Photon noise is the square root of the total number of photons collected over the 18 orbits of observations. The KLIP throughput uncertainty is obtained using the standard deviation of the percentage change in injected PSF flux when injection-recovery tests are performed at the separation of PDS 70 b at varying position angles. We find that speckle noise is the dominant component of the total error budget. It is more than 50 times greater than the photon noise and $\sim$4 times greater than the KLIP throughput uncertainty. We assume that these three components are independent and combine them in quadrature to obtain the total uncertainty in our photometry. Our final measurement of the H$\alpha$ flux of PDS 70 b is $(1.7 \pm 0.3) \times10^{-15}$ erg s$^{-1}$ cm$^{-2}$.

\section{Optimizing RDI for Planet Detection}
\label{contrast}
We explore different implementations of RDI to find the most optimal setups for planet detections. We first discuss in detail the procedure to generate contrast curves for PSF-subtracted images. We then use contrast curves to (1) determine the best combination of reference stars to use for optimal RDI subtraction, given a library size; and (2) investigate the optimal size of the reference library to be selected when performing RDI.

\subsection{Contrast Curves}
\label{contrast-curve-section}
\begin{figure}[t]
    \centering
    \includegraphics[scale=0.28]{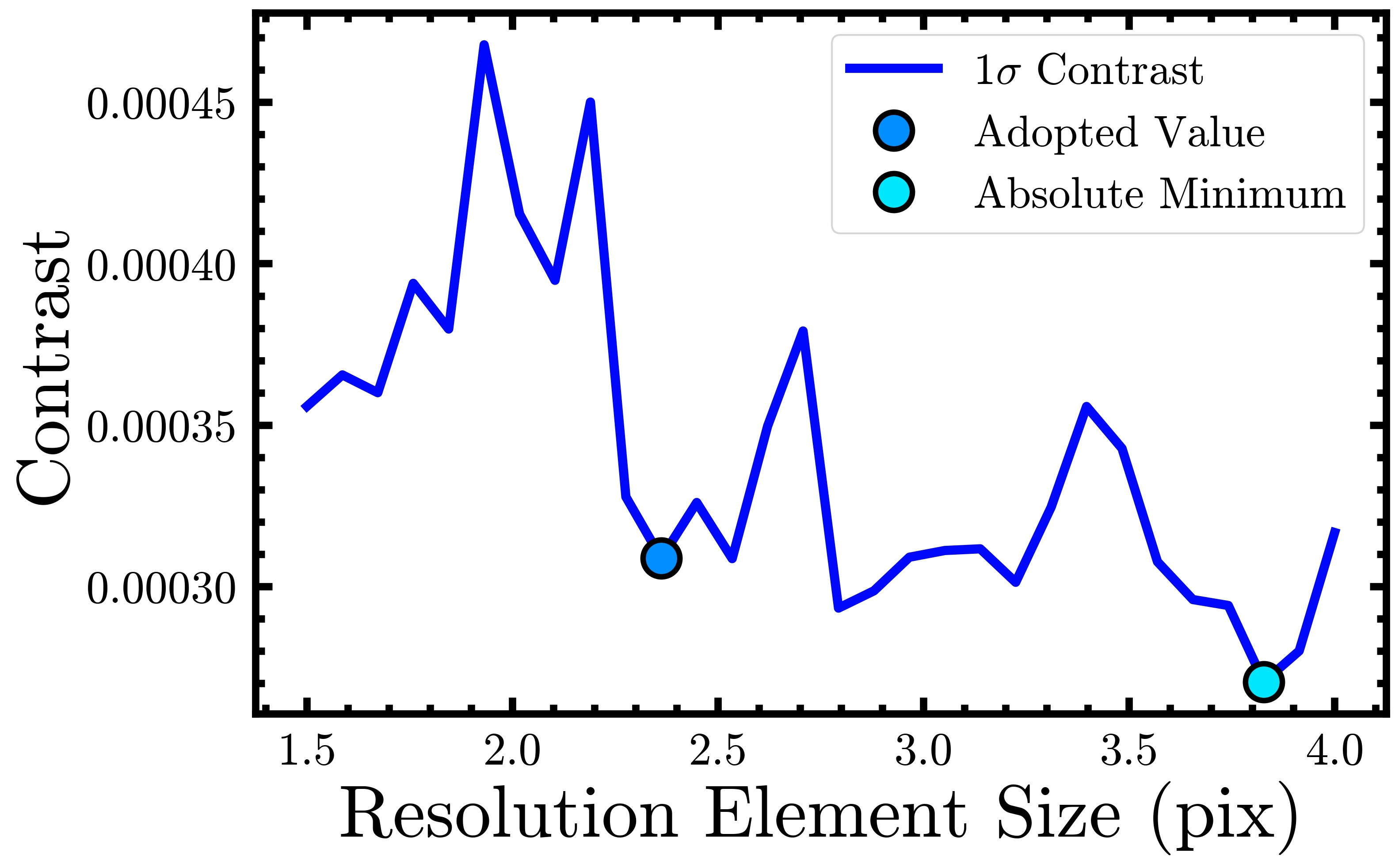}
    \caption{1$\sigma$ contrast at the separation of PDS 70 b as a function of resolution element radius. For small radii, significant variation in intensity among the resolution elements increases the 1$\sigma$ contrast level. While we get the lowest contrast level (absolute minimum) at a radius of $\sim$3.8 pixels, the number of resolution elements are too small to enable accurate computation. Balancing the two factors, we select the local minimum at 2.36 pixels and adopt this as the radius of our circular apertures for contrast curve determinations.}
    \label{fig:element-size}
\end{figure}

\citet{2014ApJ...792...97M} showed that when performing high-contrast imaging at small angular separations, the significant decrease in the number of available resolution elements dramatically affects the detection thresholds and the corresponding achievable confidence levels. At the separation scale in our images, we must account for small-sample statistics in order to obtain accurate confidence levels. We therefore follow the methodology in \citet{2014ApJ...792...97M} to compute 3$\sigma$ and 5$\sigma$ contrast curves. This is done by calculating the mean and standard deviation of the intensities across $(\frac{2\pi r}{a} - 1)$ circular apertures, where $a$ is the aperture diameter, at different separations $r$ from the host star. At the separation of PDS 70 b, the resolution element containing the planetary flux is omitted to prevent biasing the contrast calculations. We note that there is a trade-off between the aperture size and the number of apertures that can be used. Small resolution element sizes result in greater variation in intensity across a single resolution element and large resolution element sizes result in a smaller number of elements that do not accurately determine the confidence levels. We empirically find the radius of our circular apertures that would minimize the variance between the aperture-integrated fluxes. By calculating the 1$\sigma$ contrast level as a function of element size, we find that a resolution element radius of 2.36 pixels ($\sim$1.4 FWHM) works best (Figure \ref{fig:element-size}).

\begin{figure*}[t]
    \centering
    \includegraphics[scale=0.8]{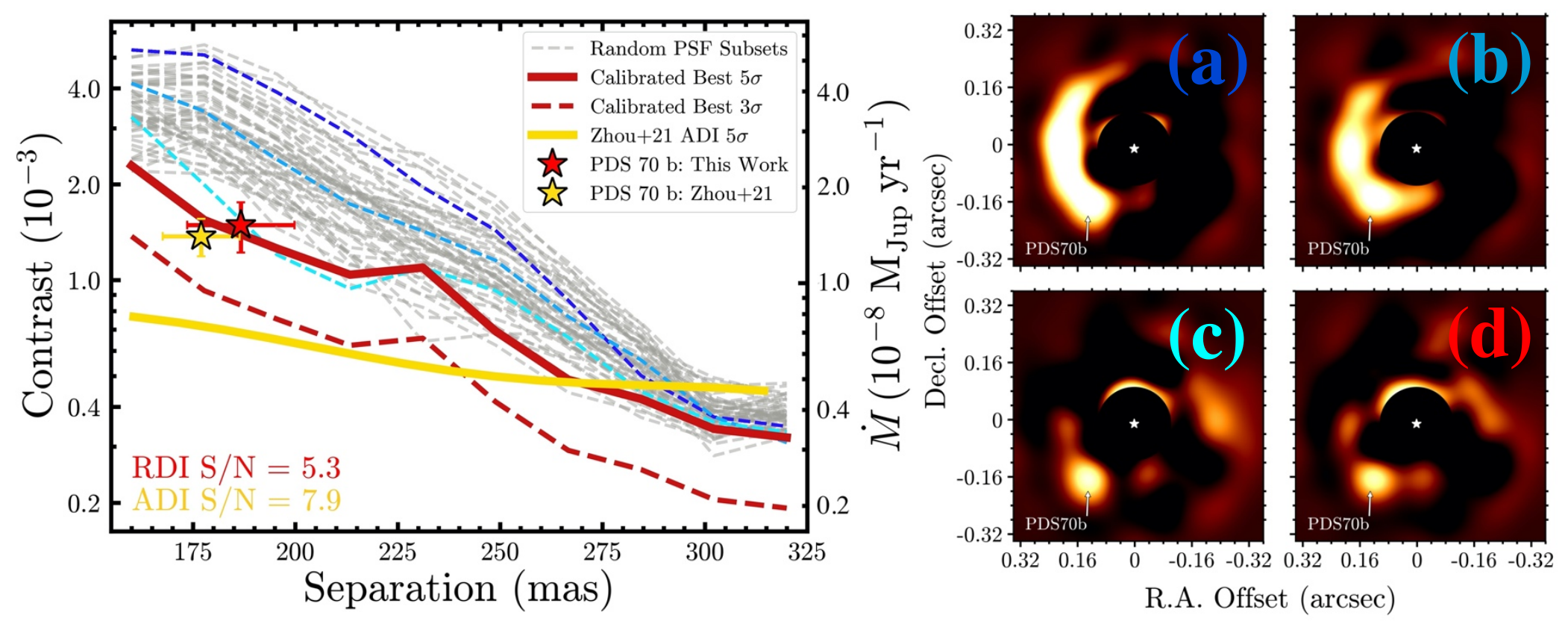}
    \caption{\emph{Left:} The distribution of 5$\sigma$ contrast curves obtained for an assortment of reference libraries consisting of 31 randomly selected reference stars, each presented as a dashed gray curve. The solid red curve is the 5$\sigma$ contrast curve and the dashed red curve is the 3$\sigma$ contrast curve for our final post-processed image shown in Figure \ref{fig:RDI}. The contrast of PDS 70 b based on our measurement of its flux is marked as a red star with the corresponding uncertainties. Our detection of PDS 70 b has a S/N of 5.3. The 5$\sigma$ contrast curve and corresponding PDS 70 b contrast from \citet{Zhou_2021} using ADI are plotted as a yellow solid line and yellow star, respectively, for comparison. Three contrast curves are selected from the distribution as examples (see below). They are colored dark blue, light blue, and cyan from the top to bottom respectively. \emph{Right:} Post-processed images corresponding to selected contrast curves in the left panel. Image (a) corresponds to the dark blue (top) contrast curve, image (b) corresponds to the light blue (center) contrast curve, image (c) corresponds to the cyan (bottom) contrast curve, and image (d) corresponds to the solid red (5$\sigma$) contrast curve. The residuals progressively reduce from images (a) to (d) at the separation of PDS 70 b and thus correspond to an improvement in detection limits.}
    \label{fig:contrast-curve}
\end{figure*}

\begin{figure*}[t]
    \centering
    \includegraphics[scale=0.07]{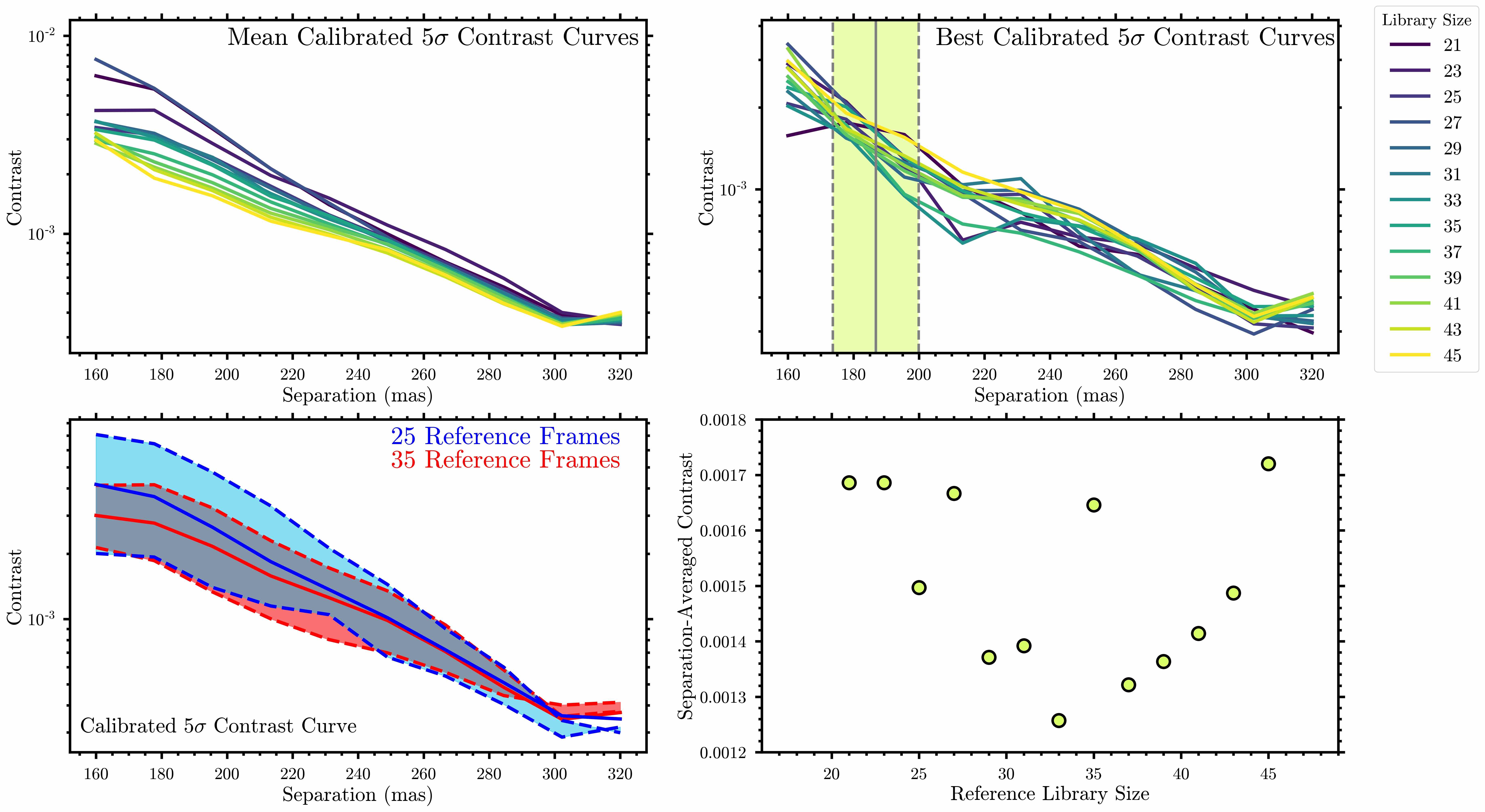}
    \caption{Assessment of the optimal size of the WFC3 reference star library for RDI reduction. \emph{Top left:} Mean of the 50 5$\sigma$ contrast curves generated for each library size depicted in the legend at the right. Brighter colors correspond to larger reference libraries. \emph{Top right:} Similar to the top left panel but presenting the best of the 50 5$\sigma$ contrast curves generated for each library size depicted in the legend. The green shaded area represents the 1$\sigma$ separation interval for our detection of PDS 70 b. 1$\sigma$ limits are illustrated by dashed gray lines and the solid line marks the center of the separation interval. \emph{Bottom left:} Example distributions of the 50 5$\sigma$ contrast curves generated for libraries of size 25 and 35. The solid lines represent the mean contrast curve, the dashed lines represent the best and the worst contrast curves for the given library size. Shaded regions highlight the spread in the contrast curve distribution from the mean. \emph{Bottom right:} Separation-averaged 5$\sigma$ contrast requirement in the shaded green region for each library size, derived using the curves in the top right panel. No correlation is found between library size and the separation-averaged 5$\sigma$ contrast requirement.}
    \label{fig:optimal-library-size}
\end{figure*}

Next, we correct the contrast curves for aperture flux loss and throughput loss. The aperture flux loss is obtained by determining the flux lost when an aperture of the size of the resolution elements is used to estimate the flux of a TinyTim PSF. We have already estimated the throughput loss as a function of separation using TinyTim PSF injection-recovery testing as discussed in Section \ref{ir-testing} and presented in Figure \ref{fig:IR}. The correction factor in each case is then obtained using the equation $\frac{1}{1 - f}$, where $f$ is the fractional flux lost in either case. The contrast curves are multiplied by the correction factors to obtain the final 3$\sigma$ and 5$\sigma$ contrast curves (Figure \ref{fig:contrast-curve}). 

To investigate the distribution of contrast curves obtained for different combinations of reference stars selected from our reference library for the purpose of RDI KLIP subtraction, we generate 5$\sigma$ contrast curves for 50 additional sets of randomly selected reference libraries (among ${45 \choose 31} = \frac{45!}{31!14!} \approx 1.7 \times 10^{11}$ possible combinations) with the requirement that Quality flag = 1 and the size of library = 31 PSF frames. The size is fixed to the number of reference stars with which we obtained our best detection of PDS 70 b to enable a direct comparison. The results are presented in Figure \ref{fig:contrast-curve}. The contrast curves vary significantly as we change the permutations of reference stars used, even though the reference stars chosen are all Quality 1 PSFs. For example, at the separation of PDS 70 b, the 5$\sigma$ contrast changes by a factor $\sim$4 between the best and the worst contrast levels. Such a large difference emphasizes an important consideration in RDI: the choice of reference stars has a dominant effect on the detection limits. We compare the 5$\sigma$ contrast curve obtained for our final processed image (solid red curve, Figure \ref{fig:contrast-curve}) with those obtained with different subsets of reference stars (dashed gray curves, Figure \ref{fig:contrast-curve}).

\subsection{Optimal Size of the Reference Library}
We quantitatively determine the best number of reference stars as well the specific reference stars to use with RDI. For a library of $N$ reference stars, there are $\sum_{k=1}^{N} {N \choose k} = 2^N - 1$ unique possible combination of images; this amounts to $2^{45} - 1 \approx 3.5 \times 10^{13}$ possible libraries for the PSF library we assembled for this study. We attempt to determine the optimal size of the reference library that would in general give us the highest S/N detection of PDS 70 b by comparing the resulting typical contrast curves. 

The compiled reference library comprises a total of 45 reference frames. We test subsets of this reference library with sizes ranging from 21 to 45 reference frames in a step size of 2. For each library size, 50 or the maximum number of allowed random configurations\footnote{For a reference library of 45 frames, there exists only ${45 \choose 45} = 1$ configuration and for a reference library of 44 frames, there exist only ${45 \choose 44} = 45$ configurations.} of the reference library are generated. RDI-KLIP image processing is conducted with each configuration and 5$\sigma$ contrast curves are generated. The mean of the 50 contrast curves is computed for each subset size and labeled as the mean uncalibrated contrast curve. Next, the mean contrast curve is calibrated for throughput and aperture loss through injection-recovery testing as described in Section \ref{contrast-curve-section}. The mean 5$\sigma$ contrast curves are presented in the top left panel of Figure \ref{fig:optimal-library-size} for each library size. Simultaneously, we also obtain the best characteristic contrast curve for each library size --- the contrast curve corresponding to the deepest 5$\sigma$ detection limits based on the 50 randomly assembled libraries for that size. Note that these may not be the absolute best contrast curves for each library size since we are sampling a small subset of possible permutations of the reference library. The best 5$\sigma$ contrast curves are presented in the top right panel of Figure \ref{fig:optimal-library-size} for each library size. Finally, we determine the separation-averaged 5$\sigma$ contrast requirement for each library size in a $\pm 1\sigma$ interval centered on the separation of PDS 70 b, using the best contrast curves. These results are presented in the bottom right panel of Figure \ref{fig:optimal-library-size}. 

The mean contrast curves show that, on average, larger library sizes result in more reliable detections; over multiple selections of the reference library, lower contrast limits are obtained with larger library sizes. However, the best characteristic contrast curves presented in the top right panel of Figure \ref{fig:optimal-library-size} demonstrate that the deepest detection limit is not necessarily obtained with the largest library. There is no correlation between the separation-averaged 5$\sigma$ contrast limit and the size of the reference library, and there are multiple sizes of the reference library that provide similarly low contrast limits at the separation of PDS 70 b. We discuss three inherent limitations to this approach:

1. This test is specific to the given set of reference stars since the library is limited by size. Only a small number of configurations are possible for reference libraries with sizes close to the maximum number of available reference frames. To fully assess the optimal library size, we need a larger reference library. Nevertheless, this test does demonstrate an approach to investigate the optimal number of reference PSFs to use for a given library.  

2. An RDI setup that improves detection sensitivity by lowering the 5$\sigma$ contrast requirement could also result in a contaminated detection of the planet due to overlapping speckles. This scenario was observed in multiple reductions when performing random selection of the reference library. The test we performed relies on contrast curves which are angularly-averaged measures and do not reflect the level of contamination in a point source detection by overlapping speckles since the resolution elements encompassing the detection are omitted in the contrast curve calculation to prevent bias.

3. Sampling every possible combination of reference images is not practically or computationally efficient. Our experiment sampled 50 libraries per size, but a larger number would likely identify even better ``best" characteristic contrast curves.

Based on the above results, the mean and best contrast curve test cannot decisively determine the best number of reference stars to use. We recommend the use of random selection to create multiple subsets of the reference library, perform RDI subtraction, and find the subset that achieves the deepest contrast limits, as indicated by their contrast curve, and gives the least contaminated point source detection, as determined by visual inspection. The random selection technique effectively samples the ensemble of reference library subsets to help eliminate contamination by speckles in reference PSFs. While it may not be possible to decisively find the best number and set of reference stars to use, it is possible to confirm whether a given detection is the best detection when the size of the reference library is fixed using the distribution of the contrast curves (Figure \ref{fig:contrast-curve}).

\begin{deluxetable*}{lccccc}
\centering
\tablecaption{Comparison of Astrometry and Photometry of PDS 70 b in $\mathrm{H}\alpha$}
\tablenum{1}

\tablehead{\colhead{Reference} & \colhead{Instrument} & \colhead{Date} & \colhead{Separation} & \colhead{P.A.} & \colhead{Line Flux}\\ 
\colhead{} & \colhead{} & \colhead{} & \colhead{(mas)} & \colhead{($^\circ$)} & \colhead{($10^{-16}$ erg s$^{-1}$ cm$^{-2}$)}} 

\startdata
\citet{Wagner_2018} & MagAO & 3 May 2018 & $183 \pm 18$ & $148.8 \pm 1.7$ & $33 \pm 18$\tablenotemark{a} \\
\citet{Wagner_2018} & MagAO & 4 May 2018 & $193 \pm 12$ & $143.4 \pm 4.2$ & $33 \pm 18$\tablenotemark{a} \\
\citet{2019NatAs...3..749H} & VLT/MUSE & 20 June 2018 & $176.8 \pm 25$ & $146.8 \pm 8.5$ & $3.9 \pm 0.7$ \\
\citet{2020AJ....159..222H} & VLT/MUSE & 20 June 2018 & $178 \pm 25$ & $147 \pm 8$ & $8.1 \pm 0.3$ \\
\citet{Zhou_2021} & HST/WFC3/UVIS & 2 February -- 3 July 2020 & $177.0 \pm 9.4$ & $143.4 \pm 3.0$ & $16.2 \pm 2.2$ \\
This Work & HST/WFC3/UVIS & 2 February -- 3 July 2020 & $186 \pm 13$ & $142 \pm 5$ & $17 \pm 3$ \\
\enddata

\tablenotetext{a}{Line flux determined by \citet{2019ApJ...885...94T} based on measurements made in \citet{Wagner_2018}.}

\label{table}
\end{deluxetable*}

\section{Discussion}
\label{discussion}
Here, we compare the results for PDS 70 b obtained using RDI with those from ADI in \citet{Zhou_2021}, examine why the RDI detection has a lower S/N than the ADI detection obtained with the same dataset, and place our work in the context of past and future implementations of RDI.

\subsection{A Comparison of ADI and RDI Astrometry and Photometry}
PDS 70 b is detected at a higher S/N of 7.9 using ADI in \citet{Zhou_2021} compared to our best RDI detection, which achieved a S/N of 5.3. We find a separation of $186 \pm 13$ mas compared to $177.0 \pm 9.4$ mas, a position angle of $142 \pm 5^\circ$ compared to $143.4 \pm 3.0^\circ$, and an H$\alpha$ flux measurement of $(1.7 \pm 0.3) \times 10^{-15}$ erg s$^{-1}$ cm$^{-2}$ compared to the $(1.62 \pm 0.22) \times 10^{-15}$ erg s$^{-1}$ cm$^{-2}$ from \citet{Zhou_2021}. All of these results are consistent within 1$\sigma$. Table \ref{table} places our astrometry and photometry measurements in the context of previous measurements in H$\alpha$.

\subsection{An Explanation for the Low S/N of the PDS 70 b Detection with RDI}
In order to understand why RDI results in a lower detection S/N than ADI, two tests are carried out: 

1. Unlike the ADI sequence of PDS 70, the reference PSFs in the archival WFC3 library are not dithered, and hence are not Fourier upsampled (see Section \ref{HCI}). To investigate whether the higher S/N achieved by ADI is because of the Fourier upsampled images reconstructed from the set of dithered PDS 70 observations, we break the dithering pattern of the PDS 70 observations by registering the science frames to a common centroid and perform ADI on the set of registered undersampled images. We find that PDS 70 b is recovered with a S/N of 7.2. \citet{Zhou_2021} performed ADI on the Fourier upsampled images and detected PDS 70 b with a S/N of 7.9. Thus, there is no significant change in the S/N between the two reductions.

\begin{figure}[b]
    \centering
    \includegraphics[scale=0.1]{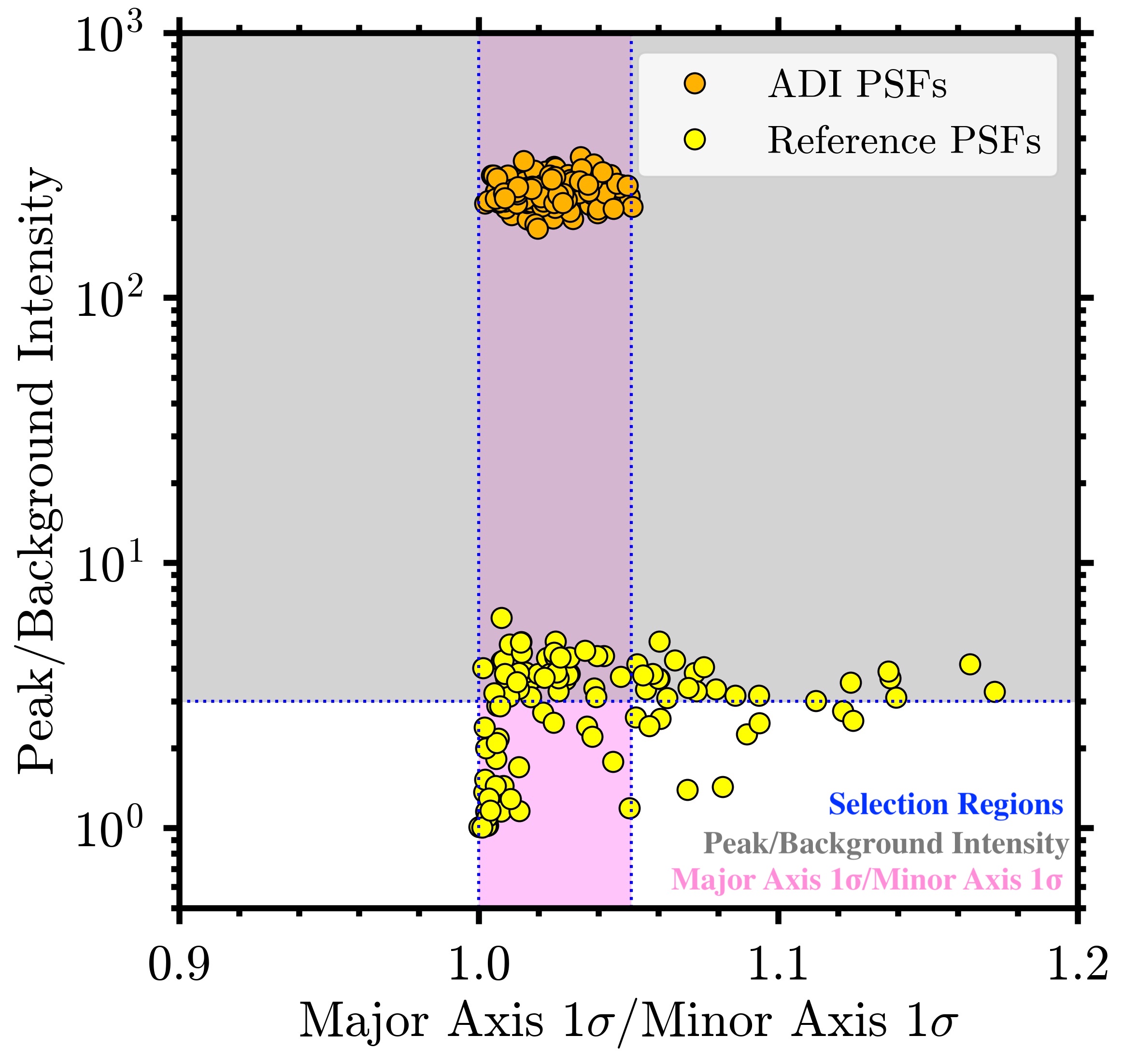}
    \caption{The morphology-significance criteria applied to the 162 ADI PSFs (orange circles). The pink and the gray regions represent the selection cuts using the morphology and significance criteria, respectively. The intersection of the two regions is the final adopted selection region. All of the ADI PSFs are in the selection region. The yellow circles show the location of our reference stars. The ADI PSFs have a peak-to-background ratio that is nearly two orders of magnitude greater than that of the reference PSFs. This highlights the significant difference in quality between the archival PSFs and the ADI PSFs.}
    \label{fig:target-selection}
\end{figure}

2. We compare the quality of our target and reference frames as quantified by the morphology-significance criteria. These selection criteria are applied to the 162 pre-processed target images as discussed in Section \ref{selection:section}. The results are presented in Figure \ref{fig:target-selection}. The peak-to-background ratios of the ADI PSFs are significantly larger by a factor of $\sim$10-100 than those of the reference stars in the library.

Since PDS 70 b is detected at a comparable S/N to the \citet{Zhou_2021} detection when ADI is performed on the undersampled images, we conclude that the dithering and image reconstruction processes are not major causes for the difference in S/N between RDI and ADI. The large difference in the peak-to-background ratios between the ADI and the reference PSFs appears to be the primary reason for the lower S/N with RDI. We expect that significant improvements can be made to the detection limits by building a reference PSF library on par with the quality of the science dataset.

\subsection{Approaches to Implementing RDI: \\The Past and the Future}
Several studies with ground- and space-based observatories have employed the RDI strategy to achieve PSF subtraction with a broad range of exoplanet- and disk-imaging science goals. Here we summarize past implementations of RDI and compare them to the strategy used in this work. We then place our methods in the context of future high-contrast imaging observations with the James Webb Space Telescope (JWST).

For ground-based observations, due to the varying observing conditions across multiple nights, a common approach is to image designated reference stars directly before and after the science observations to ensure maximum PSF correlation \citep[e.g.,][]{2017AJ....154...73R}. In the absence of designated reference star observations, \citet{2018AJ....156..156X} found that restricting the PSF library to all reference frames imaged in the same night as the science target can improve sensitivity to point sources at small angular separations with respect to ADI. \citet{2019AJ....157..118R} applied RDI using 10 reference stars with a total of $\sim$300 frames to Keck/NIRC2 observations of the low mass stellar companions HIP 79124 C and HIP 78233 B and demonstrated an increase in detection significance compared to ADI by a factor of 5. Furthermore, they showed that pre-selection of reference frames using statistical image similarity metrics such as mean-squared error, Pearson’s correlation coefficient, and structural similarity index improves detection significance by a factor of 3. Several studies have also experimented with using databases of archival PSF observations across multiple nights to build reference libraries \citep[e.g.,][]{2012SPIE.8447E..0OM, 2018MNRAS.473.1774L, 2019A&A...622A..96C, 2020AJ....159..251D}. Selection techniques involve excluding frames of close-binaries, selecting frames observed in the same spectral filter or observing mode as the science frames, and selecting frames with the highest calculated correlation coefficient between the reference and science datasets. 

Space-based applications of RDI benefit from more stable conditions compared ground-based AO. These have almost exclusively been conducted with HST, specifically with NICMOS in the IR and STIS at UV and optical wavelengths. HST PSF-subtracted images obtained using imperfectly matched PSF reference templates can suffer degradation in quality due to the introduction of optical artifacts. These artifacts primarily arise from RDI's sensitivity to chromatic differences between the science and reference frames, particularly under the very broad STIS unfiltered bandpass ($115 - 1030$ nm), and from variations in the telescope focus due to thermal instabilities known as the ``breathing" effect. To mitigate chromatic effects, reference stars are generally chosen such that their color difference with the science target is minimal \citep[e.g.,][]{2010ApJ...719.1565G, 2014ApJ...786L..23S, 2014AJ....148...59S}. To minimize breathing effects, common observing strategies include interleaving PSF reference observations with the science target in sequential, non-interruptible orbits and selecting reference stars close to the science target, typically within $10^\circ$, to reduce the spacecraft slew distance \citep{2014AJ....148...59S, 2019JATIS...5c5003D}. \citet{2021MNRAS.504.3074W} experimented with the image similarity metric PSF frame selection approach presented in \citet{2019AJ....157..118R} to select reference frames from a set of three designated reference stars for STIS observations of three protoplanetary disks. They find that the structural similarity index better discriminates between references than Pearson's correlation coefficient. Additionally, they expanded the reference library to include other STIS observations of M-type stars, but find that the results with the expanded library were of poorer quality compared to their prior results with the designated reference stars.

This work broadly differs from prior approaches in four main aspects:

1. Several of the above RDI implementations use reference frames observed in the same campaign as the science target, while ours adopts a pre-compiled PSF library not optimally built for high-contrast imaging reduction.

2. Since our science observations are restricted to the $\mathrm{H}\alpha$ narrow band (F656N) images, we avoid the complications associated with color differences between the science target and reference stars by selecting images taken in the same filter.

3. PSF selection using the morphology-significance criteria aims to quantify the quality of the reference PSFs and automatically filter out contaminated and low S/N frames. This approach fundamentally differs from previous approaches that seek to maximize the correlation between the science target and reference frames. We adopted this strategy because space-based PSFs are intrinsically more consistent than atmospheric turbulence-disturbed ground-based PSFs. Moreover, the low computing cost associated with applying the morphology-significance criteria will enable the use of much larger PSF libraries in the future.

4. Nearly all of the previous studies use coronagraphic PSFs, whereas this work uses non-coronagraphic PSFs. Previous space-based coronagraphic imaging has primarily  targeted circumstellar, protoplanetary, and transition disks at large inner working angles. For example, HST/STIS offers broadband visible-light coronagraphy using two wedge-shaped occulters
(WEDGE-A and WEDGE-B) with a minimum
half-width of $\sim$0.3$"$ and a $3"$-wide rectangular bar-shaped occulter (BAR10) \citep{2017stis.rept....3S}. Recently, STScI made available the BAR5 occulter, which has a narrow half-width of 0.15", for STIS observations. \citet{2019JATIS...5c5003D} demonstrate that better than $10^{-5}$ contrast can be achieved at angular separations $>$0.25$"$ with BAR5. We note that the morphology-significance criteria are not applicable to such coronagraphic PSF observations since these criteria are based on characterization of the PSF core. Alternative frame-correlation based techniques must be explored for PSF selection in such cases.

We anticipate the applicability of the morphology-significance criteria as a PSF selection technique for future implementations of RDI with JWST. Since the maximum allowable roll angle with JWST is $\pm 7^\circ$, RDI may be the optimal strategy for high-contrast imaging at small angular separations. Specifically, non-coronagraphic observations are expected be favorable due to the large inner working angles of the NIRCam ($0.4"-0.81"$; round mask) and MIRI ($0.33"-0.49"$; 4-quadrant phase-mask) coronagraphs. For observations that do not require very small inner working angles, coronagraphic imaging is likely preferred. Thus, for narrow-band, non-coronagraphic JWST high-contrast imaging observations, the morphology-significance criteria should provide an efficient, automated, and uniform method to select reference frames for optimal RDI subtraction. 

\section{Conclusion}
\label{conclusion}
This work has demonstrated that reference star differential imaging using F656N ($\mathrm{H}\alpha$) WFC3/UVIS images of PDS 70, in conjuction with the WFC3 archival PSF library, enables the detection of the accreting protoplanet PDS 70 b. Our results are summarized below:

1. We compiled a reference point spread function library consisting of 112 $\mathrm{H}\alpha$ reference frames in the same subarray as our observations of PDS 70 from the WFC3 archival PSF library provided by STScI. Based on visual inspection, this library consists of 45 good quality PSFs and a total of 67 sub-optimal PSFs.

2. We developed morphology-significance criteria for robust pre-selection of reference stars to improve RDI subtraction. The morphology-significance criteria automatically selects reference PSFs from the WFC3 library with an 84\% true positive fraction and 13\% false positive fraction. Combining this criteria with STScI's \texttt{QFIT} parameter enables a reduction in the number of false positives in PSF selection from 31\% when only using \texttt{QFIT} to 6\% when \texttt{QFIT} is implemented in conjunction with the morphology-significance criteria. To achieve higher true positive fractions in PSF selection, we recommend the use of the morphology-significance criteria, alone, for small reference libraries where false positives can be filtered out by visual inspection. In contrast, to reduce false positive fractions, we recommend combining the morphology-significance criteria with STScI's \texttt{QFIT} parameter for large reference libraries, where it may be time-consuming to filter out false positives by visual inspection.

3. The PSF library is used to perform reference star differential imaging on 162 $\mathrm{H}\alpha$ WFC3 images of PDS 70 and detect the planet PDS 70 b at a S/N of 5.3. The contrast curve for the detection is computed accounting for aperture loss and throughput loss using TinyTim PSF injection-recovery tests. We calibrated the astrometry and photometry by forward modeling our detection of PDS 70 b using TinyTim PSFs and obtained a separation of $186 \pm 13$ mas, a position angle of $142 \pm 5^\circ$, and a flux of $(1.7 \pm 0.3)\times10^{-15}$ erg s$^{-1}$ cm$^{-2}$. The astrometric and photometric measurements agree with the results published in \citet{Zhou_2021}. 

4. Various RDI configurations were explored to determine the optimal reference library size for RDI subtraction. Based on this dataset, we recommend the use of random PSF image selection to create multiple permutations of the reference library, perform RDI subtraction, and find the specific library that achieves the deepest contrast limits. The random selection technique effectively samples the ensemble of reference library subsets to to help eliminate the potential for speckles in reference PSFs to contaminate the astrophysical signals.

5. The reference PSFs have a lower peak-to-background ratio than the ADI PSFs by nearly two orders of magnitude. This explains the comparatively low S/N of the planetary detection of PDS 70 b with RDI. Based on these results, ADI is likely to be the better strategy for studying individual systems with HST if the goal is purely to maximize point source S/N. However, the associated strict roll angle requirements mean that ADI may not be the favorable approach for short-term (few hours to days) variability studies because of scheduling constraints for visits.

RDI with WFC3 is a powerful observing strategy for high-contrast exoplanet imaging. It has the capability to enable surveys to observe a larger number of targets more efficiently and with faster scheduling cadence than ADI by eliminating the need to observe individual targets at different telescope roll angles. We conclude that building a high S/N reference PSF library will provide unique opportunities to study short-term accretion processes and survey young stars to search for accreting protoplanets in $\mathrm{H}\alpha$ with WFC3.

\section*{Acknowledgements}
We thank the referee for a prompt report. This work is based on observations made with the NASA/ESA Hubble Space Telescope, obtained in GO program 15830 at the Space Telescope Science Institute. Support for Program number 15830 was provided by NASA through a grant from the Space Telescope Science Institute, which is operated by the Association of Universities for Research in Astronomy, Incorporated, under NASA contract NAS5-26555. We make use of observations made with the NASA/ESA Hubble Space Telescope, and obtained from the Hubble Legacy Archive, which is a collaboration between the Space Telescope Science Institute (STScI/NASA), the Space Telescope European Coordinating Facility (ST-ECF/ESA) and the Canadian Astronomy Data Centre (CADC/NRC/CSA). Y.Z. acknowledges support from the Harlan J. Smith McDonald Observatory fellowship and Heising-Simon Foundation. B.P.B. acknowledges support from the National Science Foundation grant AST-1909209 and NASA Exoplanet Research Program grant 20-XRP20$\_$2-0119.

\emph{Additional Software/Resources:} \texttt{Astropy} \citep{2013A&A...558A..33A, 2018AJ....156..123A}, \texttt{Photoutils} \citep{2019zndo...3568287B}, \texttt{Matplotlib} \citep{2007CSE.....9...90H}, \texttt{NumPy} \citep{harris2020array} \texttt{scikit-image} \citep{2014arXiv1407.6245V}, and the NASA Astrophysics Data System (ADS).

\appendix
\restartappendixnumbering
\section{Forward Modeling the PDS 70 b Detection with an ADI Template PSF}

TinyTim PSFs do not model every effect on the observed PSF and thus may not accurately represent the HST instrumental PSF. To validate our results obtained with the TinyTim PSF in Sections \ref{fm_section} and \ref{ir-testing}, we also attempted to forward model our detection of PDS 70 b with an observational template PSF.

1. The observational template PSF we make use of is the mean of the registered ADI PSFs of PDS 70 and is cropped to a $5 \times 5$ pixel frame for injection. 

2. We follow the procedures outlined in Section \ref{fm_section} to measure the astrometry and calibrate the flux of PDS 70 b by forward modeling our detection with the mean ADI PSF. The corresponding uncertainties are derived using injection-recovery tests as detailed in Section \ref{ir-testing}.

\begin{figure*}[t]
    \centering
    \includegraphics[scale=1.0]{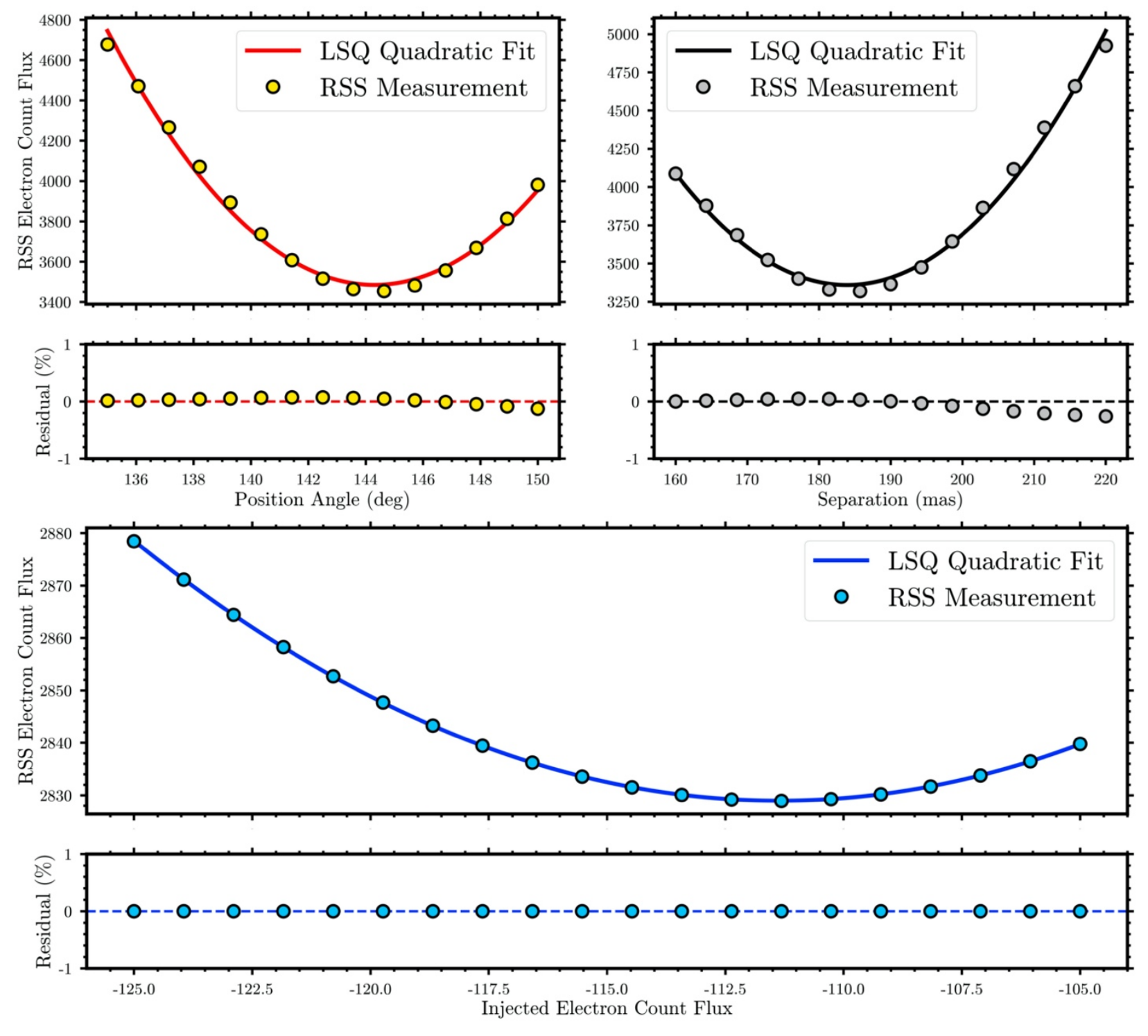}
    \caption{Calibrating astrometry and photometry of our detection of PDS 70 b with forward modeling. \emph{Top panel:} RSS electron count flux as a function of position angle and separation based on a quadratic least-squares (LSQ) fit to our measurements. Sub-panels show the fit residuals as a percentage of the RSS electron count flux for each measurement. A P.A. of $144^\circ$ with respect to north and a separation of $184$ mas from the host star minimizes the residuals. \emph{Bottom Panel:} RSS electron count flux as a function of injected electron count flux. The sub-panel shows the fit residuals as a percentage of the RSS electron count flux for each measurement. A total absolute electron count flux of 111.3 electron counts or, equivalently, $1.6\times10^{-15}$ erg s$^{-1}$ cm$^{-2}$, minimizes the residuals based on a quadratic least-squares (LSQ) fit to the measurements.}
    \label{fig:FM-adi}
\end{figure*}

\begin{figure*}[t]
    \centering
    \includegraphics[scale=0.14]{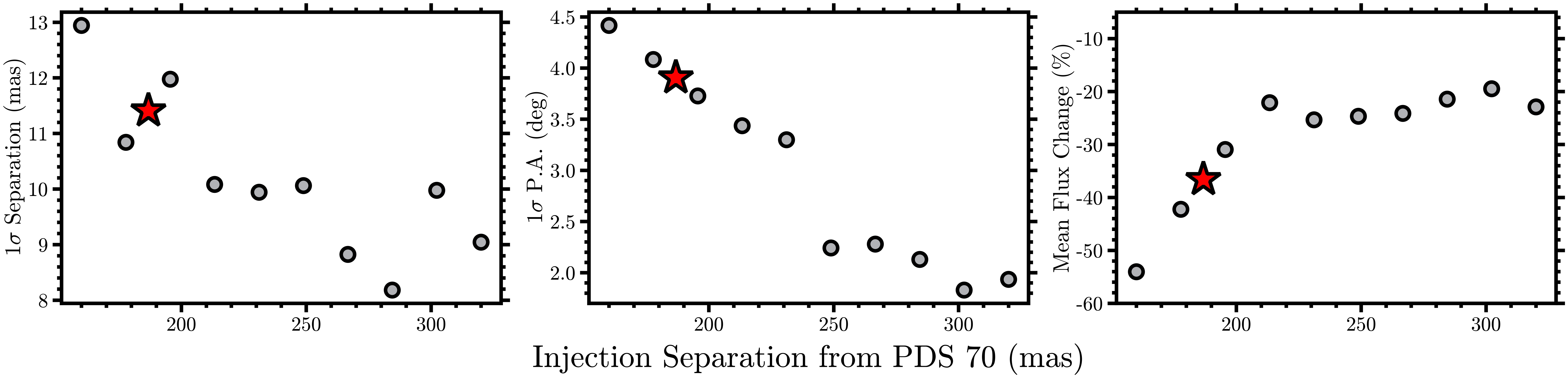}
    \caption{Error estimated by injection-recovery tests. The mean ADI template PSF is injected with the same positive flux as that obtained for PDS 70 b from forward modeling. Left: $1\sigma$ uncertainty in separation as a function of injection separation from PDS 70. Center: $1\sigma$ uncertainty in P.A. as a function of injection separation from PDS 70. Right: Mean flux change percentage as a function of injection separation from PDS 70. The astrometric uncertainty and mean flux change percentage at the separation of PDS 70 b (red stars) are calculated as the mean value of adjacent measurements.}
    \label{fig:IR-ADI}
\end{figure*}

3. Figures \ref{fig:FM-adi} and \ref{fig:IR-ADI} depict our results with the mean ADI PSF. We obtain a separation of $184 \pm 12$ mas, a P.A. measurement of $144 \pm 4^\circ$, and H$\alpha$ flux of $(1.6 \pm 0.3)\times10^{-15}$ erg s$^{-1}$ cm$^{-2}$. These values are in good agreement with those obtained using the TinyTim PSF (Table \ref{table}).

This test shows that results with the TinyTim PSFs are consistent with results obtained with the mean ADI instrumental PSF of PDS 70 and that no significant biases are introduced in our analyses due to the use of TinyTim model PSFs.

\bibliography{\string Sanghi_Astro_Bib}

\end{document}